\documentclass[12pt,a4paper,final]{iopart}
\usepackage{amscd,amssymb}
\usepackage{bm}
\usepackage{epsfig}
\usepackage{graphicx,xspace}
\usepackage{color}
\definecolor{labelkey}{cmyk}{.4,.2,0,0}

\begin{document}

\title[Off-diagonal entries of Wigner $K-$matrix ]
{Statistics of off-diagonal entries of Wigner $K-$matrix for chaotic wave systems with absorption}
\author{ Sirio Belga Fedeli and Yan V Fyodorov}
\address{King's College London, Department of Mathematics, London  WC2R 2LS, United Kingdom}

\date{\today}

\vspace{2.25cm}
\begin{abstract}
Using the Random Matrix Theory approach we derive explicit distributions of the real and imaginary parts for off-diagonal entries of the Wigner reaction matrix $\mathbf{K}$ for wave chaotic scattering in systems with and without time-reversal invariance,  in the presence of an arbitrary uniform absorption.
\end{abstract}

\maketitle

\section{Introduction}
The phenomenon of chaotic resonance scattering of quantum waves (or their classical analogues) has attracted considerable theoretical and experimental interest for the last three decades, see e.g. articles in \cite{chao05} and recent reviews \cite{kuhl13,grad14,diet15,hcao15}. The resonances manifest themselves via fluctuating structures in scattering observables, and understanding their statistical properties as completely as possible remains an important task. The main object in such an approach is the energy-dependent $M\times M$ random unitary scattering matrix $S(\lambda)$, $S^{\dagger}(\lambda)S(\lambda)={\bf 1}_M$ which relates amplitudes of incoming and outgoing waves at spectral parameter (energy) $\lambda$. Here the integer $M$ stands for the number of open channels at a given energy $\lambda$, the dagger denotes the Hermitian conjugation and ${\bf 1}_M$ is the $M\times M$ identity matrix. Statistics of fluctuations of the scattering observables over an energy interval comparable with a typical separation between resonances can be most successfully achieved in the framework of the so called 'Heidelberg approach' going back to the pioneering work \cite{Ver85}, and reviewed from different perspectives in \cite{MRW2010},  \cite{FSav11} and \cite{Schomerus2015}. In such an approach the resonance part of the $S$-matrix is expressed via the Cayley transform in terms of the resolvent of a Hamiltonian $\mathbf{H}$
representing the closed counterpart of the scattering system as
\begin{equation}\label{KW}
\mathbf{S}(\lambda)=\frac{{\bf 1}_M-i\mathbf{K}}{ {\bf 1}_M+i\mathbf{K}},\quad \hbox{with}\quad\mathbf{K}=
\mathbf{W}^{\dagger}\frac{1}{\lambda{\bf 1}_{N}-\mathbf{H}_N}\mathbf{W},
\end{equation}
where $\mathbf{W}$ is the  $N\times M$ matrix  containing the couplings between the channels and the system. For $\lambda\in\mathbb{R}$, the unitarity of $\mathbf{S}(\lambda)$ follows from Hermiticity of the Hamiltonian  represented in the framework of the Heidelberg approach by $N\times N$ self-adjoint matrix $\mathbf{H}_N$. The resulting $M\times M$  matrix $\mathbf{K}$ is known in the literature as the Wigner reaction $K-$ matrix.

To study  fluctuations induced by chaotic wave scattering one then follows the paradigm of relying upon the well-documented random matrix properties of the underlying Hamiltonian operator $\mathbf{H}$  describing quantum or wave chaotic behaviour of the closed counterpart of the scattering system. Within that approach one proceeds with replacing $\mathbf{H}_N$ with a random matrix taken from one of the classical ensembles:  Gaussian Unitary Ensemble (GUE, $\beta=2$), if one is interested in the systems with broken time reversal invariance or Gaussian Orthogonal Ensemble (GOE, $\beta=1$), if such invariance is preserved and no further geometric symmetries are present in the system.  The columns $\bm{w}_a, \, \, a=1,..,M$ of the coupling matrix  $\mathbf{W}$ can be considered either as fixed orthogonal vectors \cite{Ver85} (complex for $\beta=2$ or real for $\beta=1$), or alternatively
 as independent Gaussian-distributed random vectors orthogonal on average \cite{SokZel89}. The results turn out to be largely insensitive to the choice of the coupling as long as inequality $M\ll N\to \infty$ holds in the calculation.
 The approach proved to be extremely successful, and quite a few scattering characteristics were thoroughly investigated in that framework in the last two decades, either by the variants of the supersymmetry method or related random matrix techniques, see e.g. early papers \cite{Ver85}, \cite{Fyo97a,Fyo05} as well as more recent results in \cite{FyoSav12,Kum13,FN2015,Kum17,FST2018}. The results of such calculations are found in general to be in good agreement with available experiments in chaotic electromagnetic resonators (``microwave billiards''), dielectric microcavities and acoustic reverberation cameras ( see reviews \cite{kuhl13,grad14,diet15,hcao15}) as well as with numerical simulations of scattering in such paradigmatic model as quantum chaotic graphs \cite{Kot00} and their experimental microwave realizations \cite{microwgraphs1,microwgraphs1a,microwgraphs2,microwgraphs3}. Note that the Wigner $K-$matrix is experimentally measurable in microwave scattering systems, as it is directly related to the systems impedance matrix \cite{Hemmadi2005,Hemmadi2006a,Hemmadi2006b}.

 One of serious challenges related to theoretical description of scattering characteristics is however related
 to the fact that experimentally measured quantities suffer from the inevitable energy losses (absorption), e.g. due to
 damping in resonator walls and other imperfections. Such losses violate unitarity of the scattering matrix and are important for
 interpretation of experiments, and considerable efforts were directed towards incorporating them into the Heidelberg approach
 \cite{Fyo05}. At the level of the model (\ref{KW}) the losses can be taken into account by allowing the spectral parameter $\lambda$ to have finite imaginary part by replacing $\lambda\rightarrow\lambda+i\alpha/N\in\mathbb{C}$ with some $\alpha>0$.
 This replacement violates Hermiticity of the  Wigner matrix $\mathbf{K}$; in particular entries of $\mathbf{K}$ become now complex even for $\beta=1$. Note that our choice of scaling of the absorption term with $N$ is to ensure access to the most interesting, difficult and experimentally relevant regime when absorption term is comparable with the mean separation between neighbouring eigenvalues of the wave-chaotic Hamiltonian $H$, the latter being
 in the chosen normalization of the order $N^{-1}$ as $N\to \infty$.  The statistics of the real and imaginary parts of the diagonal entries ${\bf K}_{aa}$ in that regime was subject of considerable
 theoretical work \cite{Fyo2004,FyoSav2004,SavSomFyo2005}  and by now well-understood  and measured experimentally with good precision  for $\beta=1$ in microwave cavities \cite{Kuhletal2003,Hemmadi2005,Hemmadi2006a,Hemmadi2006b} and graphs \cite{microwgraphs1,microwgraphs1a,microwgraphs2,microwgraphs3}. Very recently first experimental results for $\beta=2$ were reported as well \cite{LawSir2018}.

 The situation with off-diagonal elements $\mathbf{K}_{a\ne b}$ is in comparison much worse. At present no theoretical results for the associated distributions are available in the literature for the most interesting and experimentally relevant case $\beta=1$ apart
 from the case of zero absorption \cite{FN2015}, and the mean value and variance for $|\mathbf{K}_{ab}|^2=(\Im \mathbf{K}_{a\ne b})^2 +(\Re \mathbf{K}_{a\ne b})^2$ \cite{RFW2004}.
  The main goal of the present paper is to fill in this gap partly by presenting the distribution of imaginary $\Im \mathbf{K}_{a\ne b}$ and real $\Re \mathbf{K}_{a\ne b}$ parts for the $K-$matrix entries in the presence of absorption:
\begin{equation}\label{defoff}
\mathbf{K}_{a,b}=\Tr\Bigg\{\Big(\left(\lambda+i\frac{\alpha}{N}\right){\bf 1}_N-\mathbf{H}_N\Big)^{-1}\bm{w}_b\otimes\bm{w}_a^T\Bigg\}
\end{equation}
  for systems with preserved ($\beta=1$) time reversal invariance, assuming the size $N\to \infty$. Our method also straightforwardly works  for the simpler case of broken time-reversal invariance ($\beta=2$) where the full distribution of $|K_{a\ne b}|^2$ has in fact already been derived for the case of uncorrelated channels, see eqs.(11)-(13) in \cite{RFW2003}. We start with briefly
  reconsidering that case in our framework, giving an alternative derivation of the result in \cite{RFW2003}, and showing that it is
simply related to the joint probability density of $\Im \mathbf{K}_{a\ne b}$ and $\Re \mathbf{K}_{a\ne b}$. We then provide a generalization of that result to the case of correlated channel vectors, see Eq.(\ref{RFC}).
 Then we concentrate on the $\beta=1$ case.



\section{Discussion of the main results}

\subsection{Systems with broken time-reversal invariance}\label{sec2}
In this case we assume that the matrix $\mathbf{H}_N$ is a complex Hermitian GUE matrix distributed according to the probabilty density $d\mathcal{P}(\mathbf{H}_N)\propto e^{-N/2\Tr\mathbf{H}_N\mathbf{H}_N^{\dag}}d\mathbf{H}_N$, so that the expectations over the ensemble are defined by  $\langle [...]\rangle_{GUE(N)}:=\int[...] d\mathcal{P}(\mathbf{H}_N)$. Initially we assume for simplicity that the components of the channel vectors $\bm{w}_a$ and $\bm{w}_b$ are uncorrelated complex gaussian random variables with mean zero and variance $1/N$, and we denote the expectation values with respect to the channel vectors by the overbar. Next, we consider the simplest nontrivial case of correlated channels characterized by a $2\times 2$ covariance matrix. Note that to compute the joint probabilty density of real and imaginary parts of $\mathbf{K}_{a,b}$ it is technically more convenient to consider its characteristic function, given by the Fourier transform
\begin{equation}\label{charfunc}
\mathcal{R}(q,q^*)=\overline{\langle\exp{i/2(q\mathbf{K}_{a,b}^{*}+q^{*}\mathbf{K}_{a,b})}\rangle}_{GUE(N)}.
\end{equation}

\subsubsection{Uncorrelated channel vectors $\bm{w}_a$ and $\bm{w}_b$.}
   The complex gaussian nature of components of the channel vectors $\bm{w}_a$  allows one to easily relate the characteristic function (\ref{charfunc}) to the GUE expectation of ratios of characteristic polynomials (for the  derivation see Sec. (\ref{A1cc}) below):
\begin{equation}\label{detGUE}
\mathcal{R}(q,q^*)=\Bigg\langle\frac{\det\left((\mathbf{H}_N-\lambda{\bf 1}_N)^2+\frac{\alpha^2}{N^2}{\bf 1}_N\right)}{\det\left((\mathbf{H}_N-\lambda{\bf 1}_N)^2+\frac{\alpha^2+|q/2|^2}{N^2}{\bf 1}_N\right)}\Bigg\rangle_{GUE(N)}
\end{equation}
The expectation values of products of ratios of characteristic polynomials (and their large$-N$ asymptotics) are well known, see
  \cite{FS03a,FS03b,BS06}. Using those results one easily obtains the following expression for $\mathcal{R}(q,q^*)\equiv \mathcal{R}(|q|)$ in the limit $N\rightarrow\infty$
  for any  value $\lambda$ of the spectral parameter belonging to the bulk of the GUE spectrum:
\begin{equation}\label{Cf}
\mathcal{R}(|q|)=\frac{|q/2|^4\exp{(-2\pi\rho(\lambda)\sqrt{\alpha^2+|q/2|^2})}}{4\alpha\sqrt{\alpha^2+|q/2|^2}}
\end{equation}
\[
\times \Bigg(\frac{\exp{(2\pi\rho(\lambda)\alpha)}}
{(\sqrt{\alpha^2+|q/2|^2}-\alpha)^2}-\frac{\exp{(-2\pi\rho(\lambda)\alpha)}}{(\sqrt{\alpha^2+|q/2|^2}+\alpha)^2}\Bigg), \quad , \lambda\in(-2,2)
\]
where $\rho(\lambda)=1/(2\pi)\sqrt{4-\lambda^2}$ is the mean density of GUE eigenvalues given by the  Wigner semicircular law.
The numerical verification of (\ref{Cf}) for systems with broken time-reversal symmetry is provided in Fig. 1 below.

Inverting this characteristic function yields the joint probability density function of $\mathbf{K}_{a,b}$ and $\mathbf{K}_{a,b}^{*}$ described in the following
\\
\vspace{0.25cm}
\noindent
\textbf{Proposition 1}:
\
\textit{Define the operator $\hat{\mathcal{D}_x}$ as}
$$
\hat{\mathcal{D}}_x=\sinh(x)\Big(1+\frac{d^2}{dx^2}\Big)-2\cosh(x)\frac{d}{dx}
$$
\textit{Then  the joint probability density function of the pair $(\mathbf{K}_{ab},\mathbf{K}_{ab}^*)$, with $\mathbf{H}_N\in GUE(N)$
in the limit $N\rightarrow \infty$  is given by:}
\begin{equation}\label{result1}
\mathcal{P}(\mathbf{K}_{a,b},\mathbf{K}_{a,b}^{*})=\frac{\alpha^2}{\pi}\lim_{x\rightarrow 2\pi\rho(\lambda)\alpha}\hat{\mathcal{D}}_x\frac{\exp{(-\sqrt{x^2+4\alpha^2|\mathbf{K}_{a,b}|^2})}}{\sqrt{x^2+4\alpha^2|\mathbf{K}_{a,b}|^2}}
\end{equation}

\begin{figure}[h!]
\centering
 {\includegraphics[width=.70\textwidth]{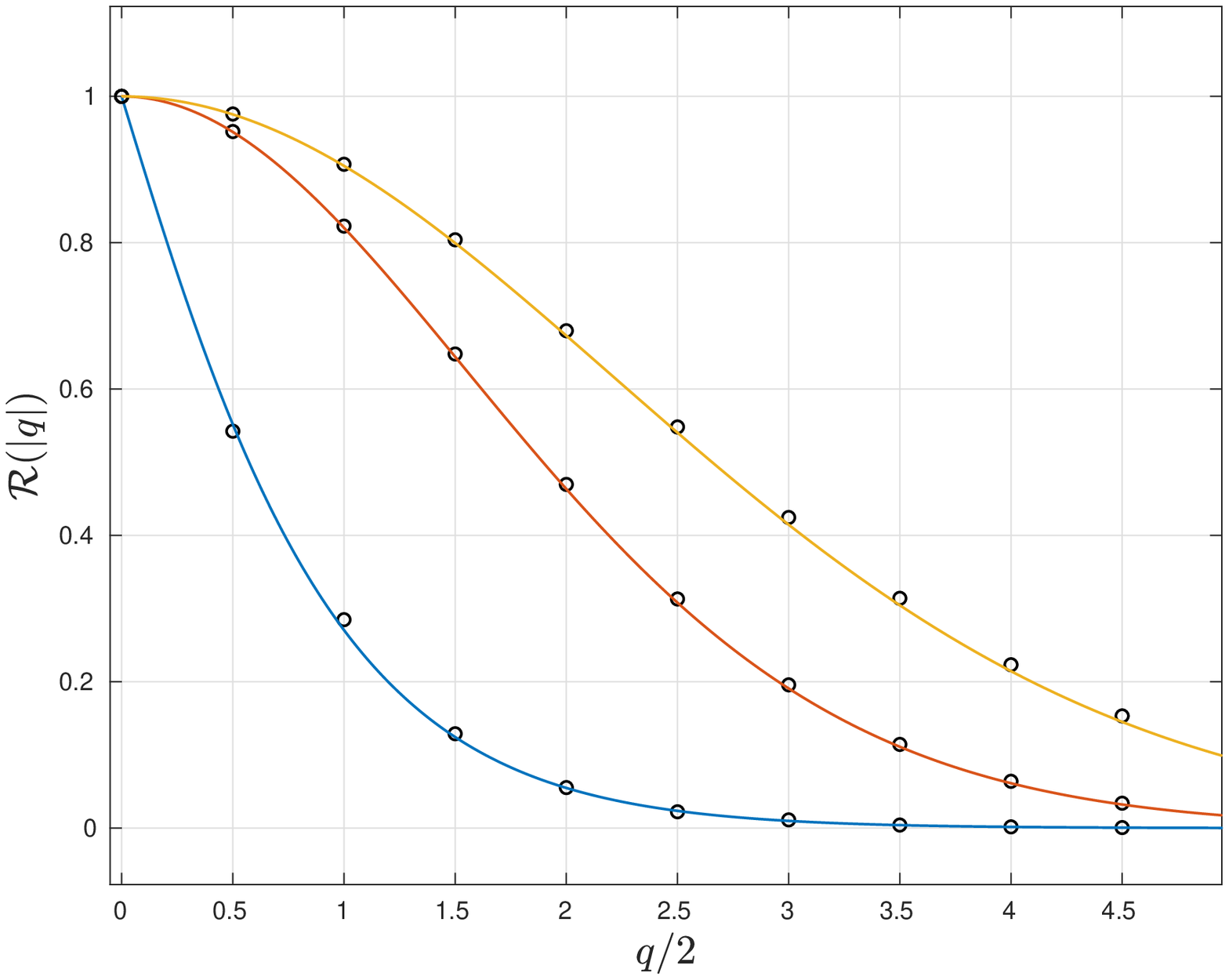}} \\
   {\includegraphics[width=.70\textwidth]{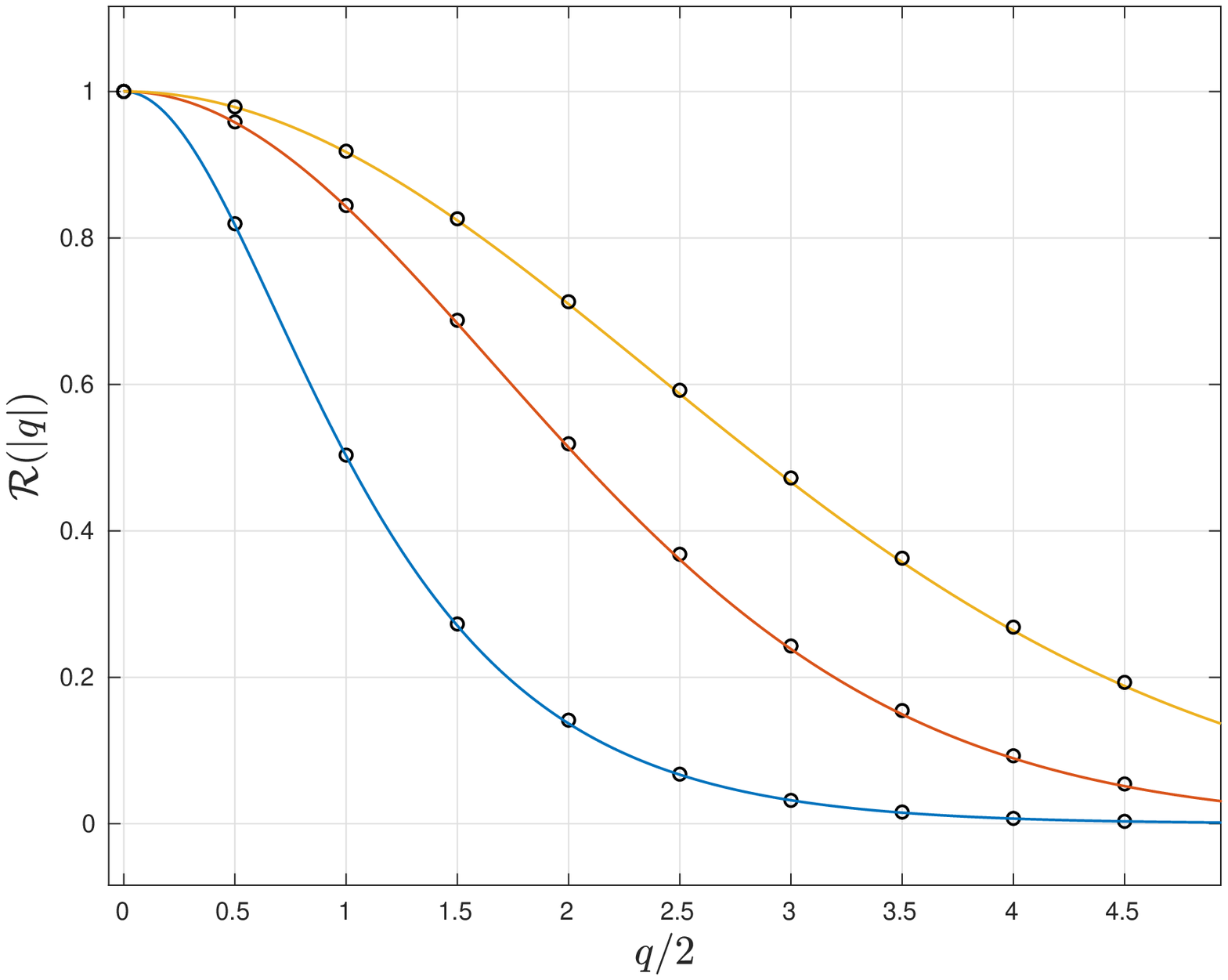}} \\
\label{pp10}
\caption{Characteristic function Eq.(\ref{Cf}) for systems with broken time-reversal invariance for $\alpha=1(blue),\,\alpha=5(red),\, \alpha=10(yellow)$
and  $\lambda=0 (top),\,\lambda=1(bottom)$ versus direct numerical simulations (\# $N=100$, 50000 samples, circular markers).}
\end{figure}

\vspace{0.25cm}
In the case $\alpha\rightarrow0^{+}$, which corresponds to vanishing absorption, the joint density $\mathcal{P}(\mathbf{K}_{a,b},\mathbf{K}_{a,b}^{*})$ acquires a simple form:
$$
\mathcal{P}\left(\Re\mathbf{K}_{a,b},\Im\mathbf{K}_{a,b}\right)=\frac{\rho(\lambda)}{4}\frac{|\mathbf{K}_{a,b}|^2+4\pi^2\rho^2(\lambda)}{(|\mathbf{K}_{a,b}|^2+\pi^2\rho^2(\lambda))^{5/2}}
$$
which implies due to the rotational symmetry that the variables  $u_1=\Re\mathbf{K}_{a,b}$ and
$u_2=\Im\mathbf{K}_{a,b}$  each are distributed, respectively, as:
$$
\mathcal{P}(u_i)=\frac{\rho(\lambda)}{2}\frac{u_i^2+3\pi^2\rho^2(\lambda)}{(u_i^2+\pi^2\rho^2(\lambda))^2}, \quad i=1,2
$$
  Note that the results for arbitrary spectral parameter $\lambda\in(-2,2)$ could be obtained from those for  $\lambda=0$ by rescaling $\alpha\rightarrow\alpha\eta$ and $|q|\rightarrow |q|\eta$ with the ratio $\eta=\rho(\lambda)/\rho(0)$. This is a particular manifestation of the well-known {\it spectral universality} of random matrix results, the property which we will use in the next section to deal with a  much  more challenging problem of scattering systems with preserved time-reversal invariance.

\subsubsection{Correlated channel vectors $\bm{w}_a$ and $\bm{w}_b$.}

To consider the simplest nontrivial correlations between channel vectors we assume that the entries $w_{a,n}$ (resp. $w_{b,n}$) with different values of $n$ remain independent and identically distributed gaussian variables, whereas $w_{a,n}$ and $w_{b,n}$ with the same value of $n$ are correlated. The correlations can be then  described by a non-diagonal $2\times2$ Hermitian positive definite covariance matrix $\mathbf{C^{-1}}$ such that  $\overline{\bm{w}_a^{\dagger}\bm{w}_b}=\left(C^{-1}\right)_{ab}$. The corresponding joint probability density $\mathcal{P}\left(
    w_{a,n},
    w_{b,n}\right)$ has the form
$$
\mathcal{P}\left(
    w_{a,n},
    w_{b,n}\right)\propto\exp\Big\{
-N\left[\begin{array}{c}
    w_{a,n} \\
    w_{b,n} \\
\end{array}\right]^{\dag}\mathbf{C}
\left[\begin{array}{c}
    w_{a,n} \\
    w_{b,n} \\
\end{array}\right]
\Big\}
$$
It is then easy to show that the resulting characteristic function is again expressed as the ratio of determinants, similar to the uncorrelated case:
\begin{equation}\label{GUEeqCoupled}
\fl \mathcal{R}(q,q^*)=
\Bigg\langle\frac{\det\left((\lambda{\bf 1}_N-\mathbf{H}_N)^2+\frac{\alpha^2}{N^2} {\bf 1}_N\right)}{\Pi_{l=1,2}\det\left((\lambda{\bf 1}_N-\mathbf{H}_N)+\frac{i}{2N}(\tilde{k}+(-1)^l\sqrt{\tilde{k}^2+4\tilde{s}}){\bf 1}_N\right)}\Bigg\rangle_{GUE(N)},
\end{equation}
where we denoted:
$$
\tilde{k}=\frac{1}{2}\Big(\frac{C_{ab}^*}{\det\mathbf{C}}q+\frac{C_{ab}}{\det\mathbf{C}}q^*\Big)
$$
$$
\tilde{s}=\alpha^2+\frac{\alpha}{2}\Big(\frac{C_{ab}^*}{\det\mathbf{C}}q-\frac{C_{ab}}{\det\mathbf{C}}q^*\Big)+\frac{|q|^2}{4\det{\mathbf{C}}}
$$
It is useful to recall that $\frac{C_{ab}}{\det\mathbf{C}}=-\left(C^{-1}\right)_{ab}$ so reflects the nonvanishing correlations between
the two channel vectors.

Note that we assume the entries of $\mathbf{C}$ to be of order unity: $C_{i,j=a,b}=O(1)$ as $N\to \infty$. Evaluating then the expectation over the GUE matrices in (\ref{GUEeqCoupled}), we find in the large-$N$ limit:
\begin{equation}\label{RFC}
\fl \mathcal{R}(q,q^*)=
\frac{e^{-\frac{1}{2}(i\tilde{k}\lambda)-\pi\rho(\lambda)(\sqrt{\tilde{k}^2+4\tilde{s}}+2\alpha)}}{8(\sqrt{\tilde{k}^2+4\tilde{s}})
\alpha}\Big((1-e^{4\pi\alpha\rho(\lambda)})\tilde{k}^2-(\sqrt{\tilde{k}^2+4\tilde{s}}-2\alpha)^2+e^{4\pi\alpha\rho(\lambda)}(\sqrt{\tilde{k}^2+4\tilde{s}}
+2\alpha)^2\Big)
\end{equation}
Examples of $\mathcal{R}(q,q^*)$ for different levels of absorption are shown in Fig. 2, Fig. 3 and Fig.4 for a particular choice of the covariance matrix
$\mathbf{C}^{-1}=
\left[\begin{array}{cc}
   2  & -i \\
   i  &1 \\
\end{array}\right]
$. 
Small discrepancies between simulations and the theoretical formula visible in the figures are due to the finite size of the matrices used in simulations and can be checked to gradually disappear as the size of the matrices is increased. The latter effect is the more  pronounced the bigger values of $\alpha$ are considered.

\begin{figure}[h!]
\centering
{\includegraphics[width=.70\textwidth]{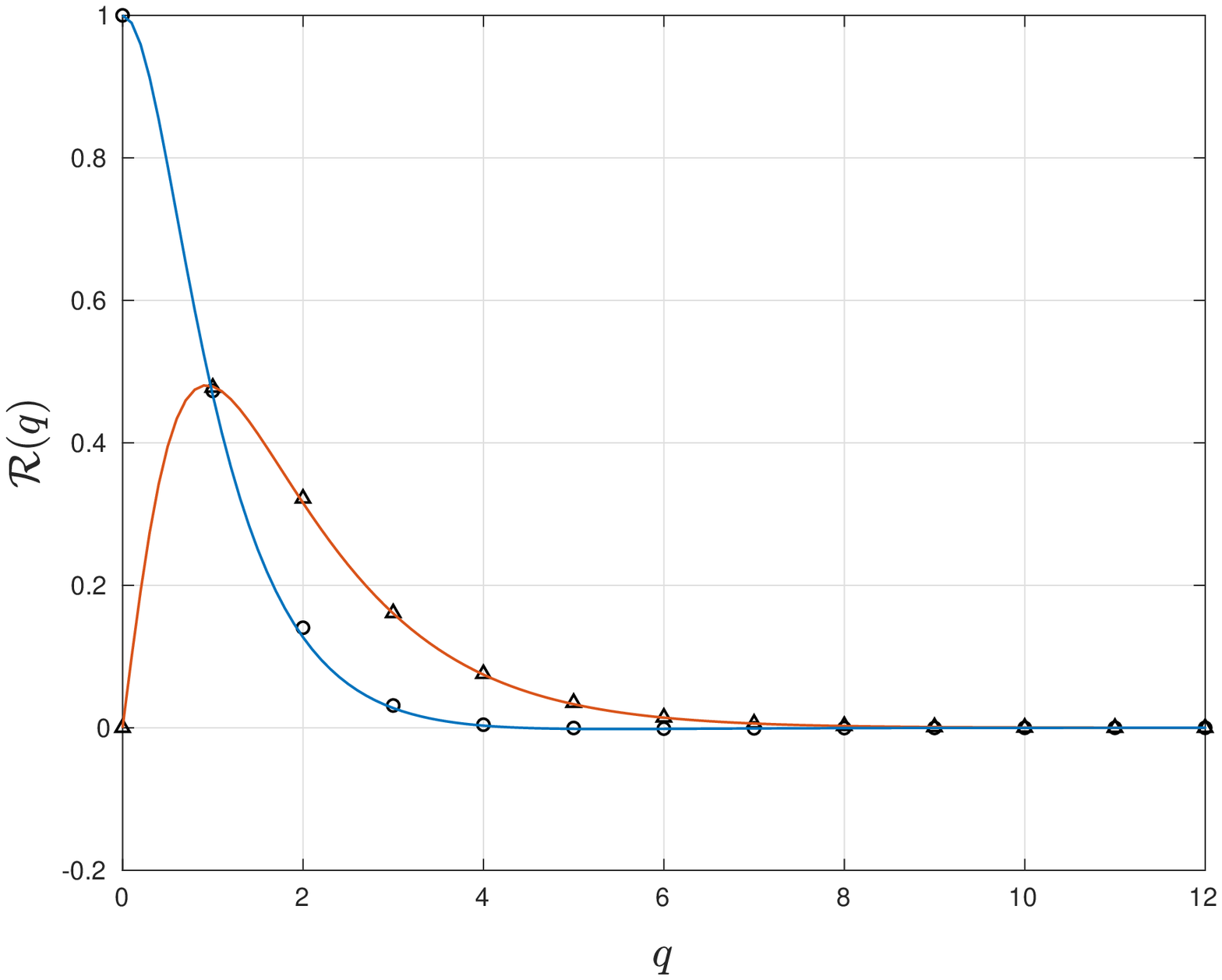}} \\
{\includegraphics[width=.70\textwidth]{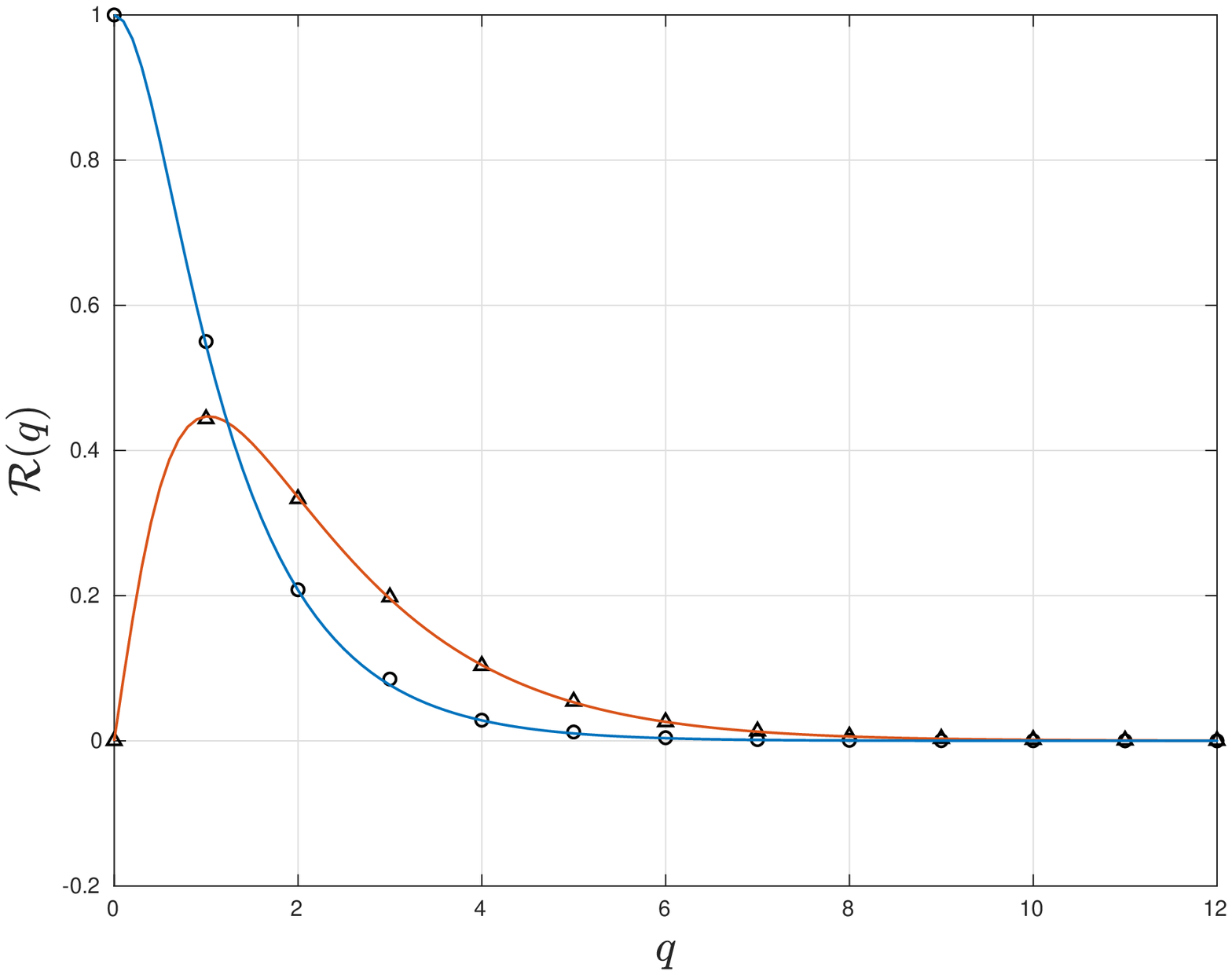}} \\
\label{Cpp10A}
\caption{Real (blue) and imaginary (red) parts of the characteristic function Eq.(\ref{RFC}) for $K_{a,b}$ in systems with broken time-reversal invariance and absorption $\alpha=1$ with the special choice of the channel covariance matrix,  for $\lambda=0$ (top),$\lambda=1$ (bottom). Markers indicate numerical results involving \# Samples 10000 for the matrix size $N$=100.}
\end{figure}
\begin{figure}[h!]
\centering
 {\includegraphics[width=.70\textwidth]{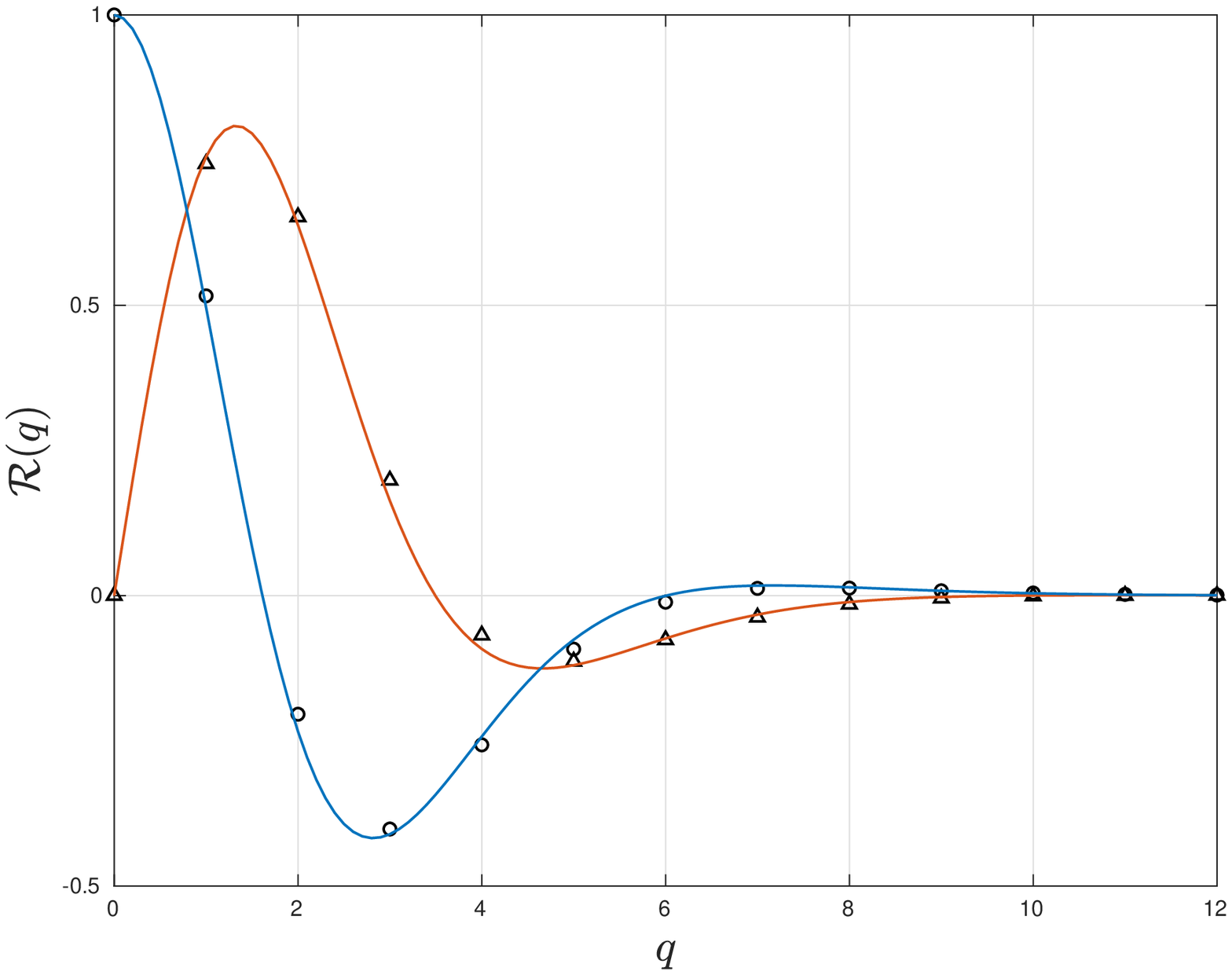}} \\
   {\includegraphics[width=.70\textwidth]{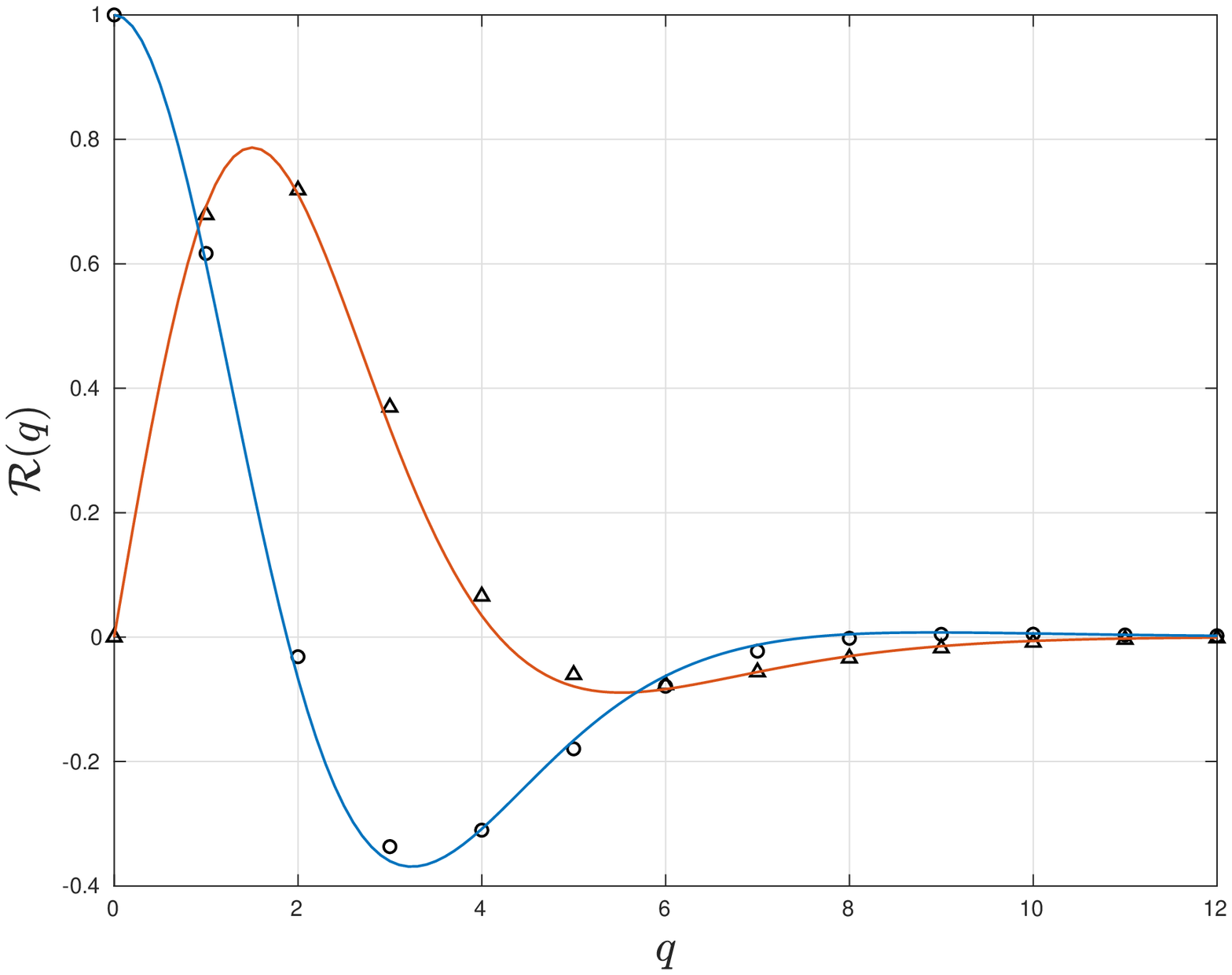}} \\
\label{CCpp10A2}
\caption{ The same as above for absorption $\alpha=5$ }
\end{figure}
\begin{figure}[h!]
\centering
 {\includegraphics[width=.70\textwidth]{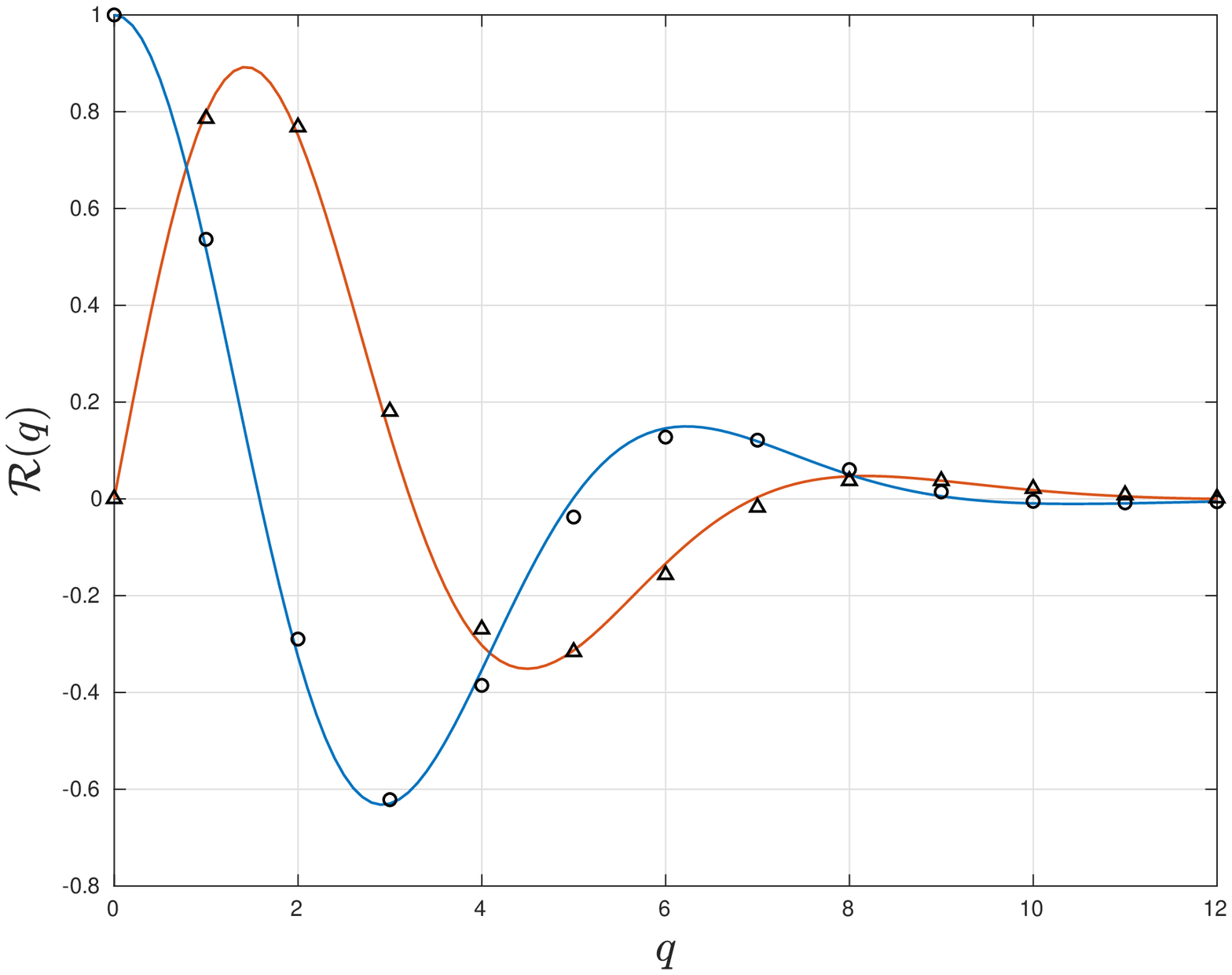}} \\
   {\includegraphics[width=.70\textwidth]{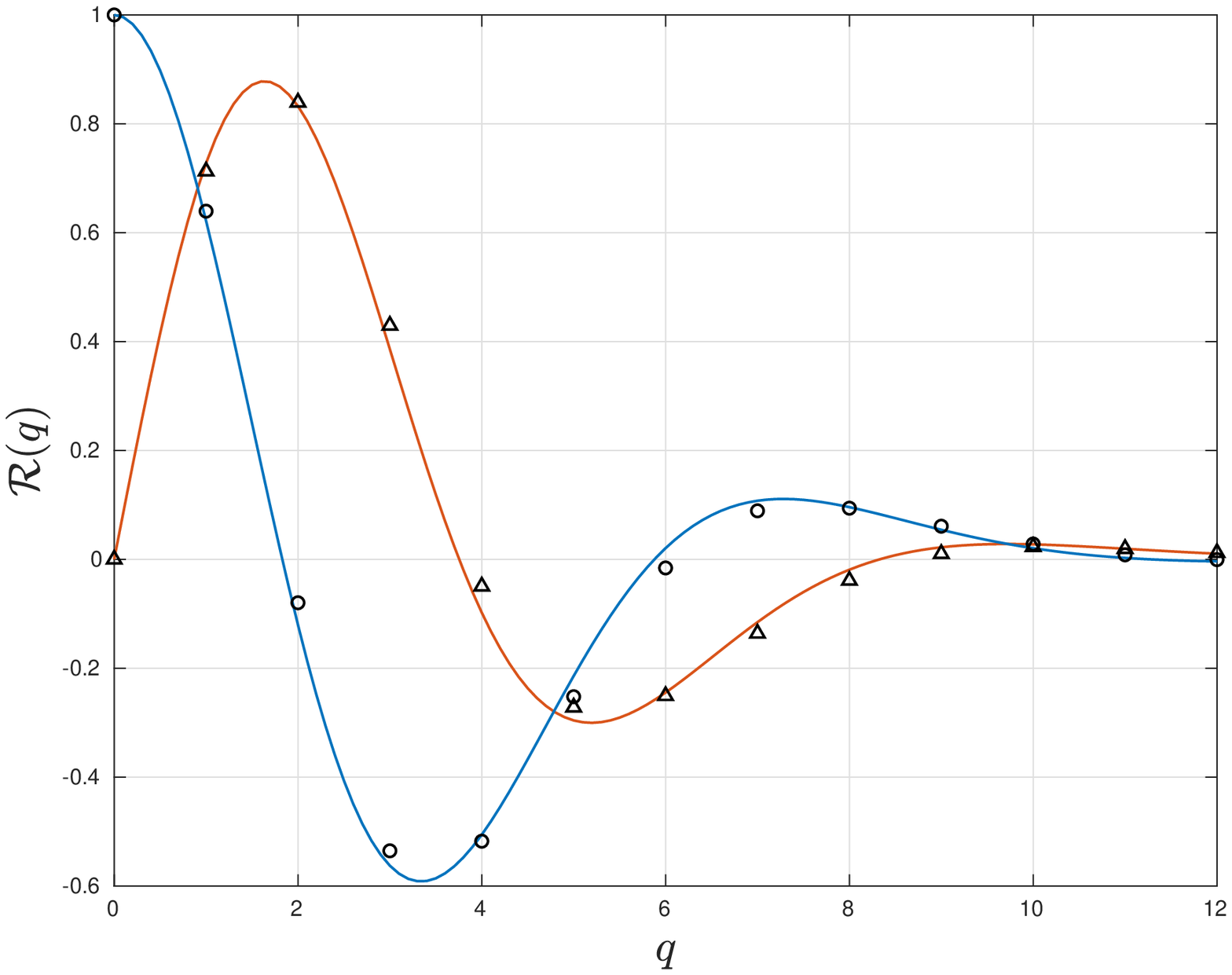}} \\
\label{CCpp10A3}
\caption{ The same as above for absorption $\alpha=10$. }
\end{figure}


\subsection{Systems with preseved time-reversal invariance}\label{sec3}
In such a case we assume $\mathbf{H}_N$ to be the real symmetric GOE matrix distributed with the probability density $d\mathcal{P}(\mathbf{H}_N)\propto\exp(-\frac{N}{4}\Tr \mathbf{H}_N^2)d\mathbf{H}_N$, whereas the channel vectors $\bm{w}_{a}$ are assumed to be independent for $a\ne b$ and their components are chosen to be real i.i.d. mean-zero gaussian random variables of variance $Var[\bm{w}_{aj}]=1/N, \,  j=1\ldots,N$. In what follows we denote the corresponding expectations with $\overline{[...]}=\int\int[...]\mathcal{P}(\bm{w}_a)\mathcal{P}(\bm{w}_b)d\bm{w}_ad\bm{w}_b$ and $\langle [...]\rangle_{GOE(N)}=\int[...]\mathcal{P}(\mathbf{H}_N)d\mathbf{H}_N$ respectively. As before, to address the distributions of  $\mathbf{K}_{a,b}=\Re \mathbf{K}_{a,b}+i \Im \mathbf{K}_{a,b}$ we introduce its characteristic function $\mathcal{R}(k,s),  q=k+is$ as in (\ref{charfunc}). Integrating out the Gaussian-distributed channel vectors one arrives at the following representation:
\begin{equation}\label{GOEeq}
\fl \mathcal{R}(k,s)=
\Bigg\langle\frac{\det\left((\lambda{\bf 1}_N-\mathbf{H}_N)^2+\frac{\alpha^2}{N^2} {\bf 1}_N\right)}{\Pi_{l=1,2}\det^{1/2}\left((\lambda{\bf 1}_N-\mathbf{H}_N)^2+(-1)^li\frac{k}{N}(\lambda{\bf 1}_N-\mathbf{H}_N)+\frac{\omega_l^2}{N^2}{\bf 1}_N\right)}\Bigg\rangle_{GOE(N)},
\end{equation}
where we denoted $\omega_1^2=\alpha^2-i\alpha s$ and $\omega_2^2=\alpha^2+i\alpha s$.

 The major difficulties in evaluating such type of GOE averages arise from the presence of half-integer powers in the denominator.
 Using a variant of the supersymmtry approach one can derive a finite$-N$ representation for the above as an integral over $4\times 4$ positive definite matrices (see Appendix A), but its asymptotic/saddle-point analysis for $N\gg 1$ presents a considerable technical challenge,
 see a detailed discussion in  \cite{FN2015}. Note also the lack of rotational invariance in the plane of the complex variable $q$
  as, in contrast to the GUE case, $\mathcal{R}(k,s)$ does not depend only on $|q|=\sqrt{k^2+s^2}$ . Athough all this prevented us from finding the full joint probability density for the pair $(\mathbf{K}_{a,b},\mathbf{K}_{a,b}^*)$, we succeeded in extracting the (most important) special cases:
 \begin{equation}
 \mathcal{R}(s,0)=\langle\overline{\exp{\left(i s\Im{\mathbf{K}_{a,b}}\right)}}\rangle_{GOE(N)}, \quad  \mathcal{R}(0,k)=\langle\overline{\exp{\left(i k\Re{\mathbf{K}_{a,b}}\right)}}\rangle_{GOE(N)}
 \end{equation}
 yielding the characteristic function for separately imaginary and real part of $\mathbf{K}_{a,b}$.
    Note that those quantities can be separately measured. To simplify our calculation we concentrate on the spectral centre $\lambda=0$, the general case $\lambda\neq 0$  recovered by using the spectral universality via rescaling with the mean spectral  density, exactly like in $\beta=2$ case.  Our main result for the characteristic functions is given by

\vspace{0.25cm}
\noindent
\textbf{Proposition 2}:
{\it Consider $\mathbf{H}_N\in GOE(N)$, $\bm{w}_{c}\sim\mathcal{N}(\bm{0},\frac{1}{N}{\bf 1}_N)$ for $c=a,b$ and define the functions
\begin{equation}
\fl C(q_1,q_2,\alpha)=q_2^2-4q_2^3+4q_1^3(4q_2-1)+2q_1q_2(1-4q_2+8q_2^2)+q_1^2(1-8q_2+44q_2^2)
\end{equation}
\[
\fl -4(q_1+q_2)\left(-q_2^3+q_1^2q_2(4q_2-5)+q_1q_2^2(4q_2-5)+q_1^3(4q_2-1)\right)\alpha^2+16q_1^2q_2^2(q_1+q_2)^2\alpha^4
\]
and
\begin{equation}
\fl D(q_1,q_2,\alpha)=C(q_1,q_2,\alpha)-8(q_1+q_2)^2\alpha^2\left(q_1+q_2-2q_1q_2+4q_1q_2(q_1+q_2)\alpha^2\right)
\end{equation}
Then  the characteristic function of the real $\Re$ and imaginary $\Im$ parts of $\mathbf{K}_{ab}$ for $\lambda=0$  are given in the limit $N\to \infty$ by the following integral representations:
\[
\lim_{N\rightarrow \infty}\overline{\Big\langle{e^{is\Im \mathbf{K}_{a,b}}}\Big\rangle}_{GOE(N)}=-\int_{\mathbb{R}_+}dq_1\int_{\mathbb{R}_+}dq_2|q_1-q_2|J_{0}\Big(s\alpha(q_1-q_2)\Big)
\]
\begin{equation}\label{Im}
\fl \times\,\, e^{-\frac{1}{4}(q_1+q_2)\left((q_1q_2)^{-1}+4\alpha^2\right)}\,\, \frac{D(q_1,q_2,\alpha)\sinh(2\alpha)-2\alpha C(q_1,q_2,\alpha)\cosh(2\alpha)}{512\sqrt{\pi}q_1^3q_2^3(q_1+q_2)^{5/2}\alpha^3}
\end{equation}
and
\[
\lim_{N\rightarrow \infty}\overline{\Big\langle{e^{ik\Re \mathbf{K}_{a,b}}}\Big\rangle}_{GOE(N)}=-\int_{\mathbb{R}_+}dq_1\int_{\mathbb{R}_+}dq_1dq_2|q_1-q_2|
I_{0}\Big(k\sqrt{\frac{k^2}{4}+\alpha^2}(q_1-q_2)\Big)
\]
\begin{equation}\label{Re}
\fl \times\,\, e^{-\frac{1}{4}(q_1+q_2)\left((q_1q_2)^{-1}+2(k^2+2\alpha^2)\right)}\,\,\frac{D(q_1,q_2,\alpha)\sinh(2\alpha)-2\alpha C(q_1,q_2,\alpha)\cosh(2\alpha)}{512\sqrt{\pi}q_1^3q_2^3(q_1+q_2)^{5/2}\alpha^3}
\end{equation}
where $J_0(x)$ and $I_0(x)$ are Bessel function and modified Bessel function, respectively.}

{\bf Note 1}. The probability density function of the scaled imaginary part $u=\alpha^{-1}\Im K_{ab}$ can be written in a closed form by inverting
the associated Fourier transform. Namely, denoting  the integrand function in equation (\ref{Im}) for $s=0$ as $f(q_1,q_2;\alpha)$ one  gets
\begin{equation}\label{P11}
\mathcal{P}(u)=\int_{0}^{1}\int_{0}^{1}dpdt f\Big(\frac{|u|}{(2pt)}(t+1),\frac{|u|}{(2pt)}(1-t);\alpha\Big)\frac{|u|}{p^2t^2\sqrt{1-p^2}}
\end{equation}
A similar inversion in (\ref{Re}) seems however impractical. \\[0.5ex]

{\bf Note 2}.  Introduce the GOE
semicircle eigenvalue density $\rho(\lambda)=1/(2\pi)\sqrt{4-\lambda^2}$ and the ratio $\eta=\rho(\lambda)/\rho(0)$.
Then characteristic functions $\langle\overline{\exp{\left(is\Im \mathbf{K}_{a,b}\right)}}\rangle_{GOE(N)}$ and $\langle\overline{\exp{\left(ik\Re \mathbf{K}_{a,b}\right)}}\rangle_{GOE(N)}$ for any $\lambda\in(-2,2)$  in the limit $N\rightarrow\infty$ can be obtained from the case $\lambda=0$ by rescalings $\alpha\rightarrow\eta\alpha$, $s\rightarrow\eta s$ and $k\rightarrow\eta k$. Namely:
\begin{equation}
\label{Rescale}
\lim_{N\rightarrow\infty}\overline{\Big\langle e^{{is\Im \mathbf{K}_{a,b}}}\Big\rangle}_{GOE(N)}(\alpha,\lambda)=\lim_{N\rightarrow\infty}\overline{\Big\langle e^{{is\eta\Im \mathbf{K}_{a,b}}}\Big\rangle}_{GOE(N)}(\eta\alpha,0)
\end{equation}
and
\begin{equation}
\label{Rescale2}
\lim_{N\rightarrow\infty}\overline{\Big\langle e^{{ik\Re \mathbf{K}_{a,b}}}\Big\rangle}_{GOE(N)}(\alpha,\lambda)=\lim_{N\rightarrow\infty}\overline{\Big\langle e^{{ik\eta\Re \mathbf{K}_{a,b}}}\Big\rangle}_{GOE(N)}(\eta\alpha,0)
\end{equation}
An analogous result holds for the probability distribution $\mathcal{P}(\cdot;\alpha,\lambda)$ in eq. (\ref{P11}) with $\lambda\neq  0$; now rescaling $\tilde{u}=\eta^2 u$, i.e. $\mathcal{P}(\tilde{u};\varepsilon,\lambda)=\mathcal{P}(\tilde{u};\eta\alpha,0)$.
The numerical comparison is presented in Figs. 2 and 3 below. A justification for the rescaling can been provided based on the results in \cite{FN2015}, see appendix A.  \\[0.5ex]

{\bf Note 3.} For large values of absorption $\alpha\gg 1$ it is natural to expect that the probability density of $\Im \mathbf{K}_{a,b}$ should approach the Gaussian shape. This is fully confirmed by Fig. 4. \\[0.5ex]

{\bf Note 4}.  Although we were not able to find the joint probability density for the pair $\Re \mathbf{K}_{a,b}$ and $\Im \mathbf{K}_{a,b}$,
we can shed some light on cross-correlations between the imaginary and real parts by using the results of the paper \cite{RFW2004} for the variance of $|\mathbf{K}_{ab}|^2=(\Im \mathbf{K}_{a\ne b})^2 +(\Re \mathbf{K}_{a\ne b})^2$.  Namely, in Fig. 5 we present the quantity  $$\tau(\alpha)=\frac{\overline{\langle\left(\Im K_{a,b}\Re K_{a,b}\right)^2\rangle}}{\overline{\langle
\left(\Im K_{a,b}\right)^2\rangle}\quad \overline{\langle\left(\Re K_{a,b}\right)^2\rangle}}-1, $$ where
$\overline{\langle\left(\Im K_{a,b}\right)^2\rangle}$ and $\overline{\langle\left(\Re K_{a,b}\right)^2\rangle}$ are calculated in the Appendix B. We see that the real and imaginary parts are correlated, but gradually decorrelate with increased absorption.

\bigskip

{\bf Acknowledgments:}
The research presented in this paper was supported by  EPSRC grant  EP/N009436/1 "The many faces of random characteristic polynomials"
and by the EPSRC Centre for Doctoral Training in Cross-Disciplinary Approaches to Non-Equilibrium
Systems (CANES, EP/L015854/1)

\begin{figure}[t]
\centering
 {\includegraphics[width=.70\textwidth]{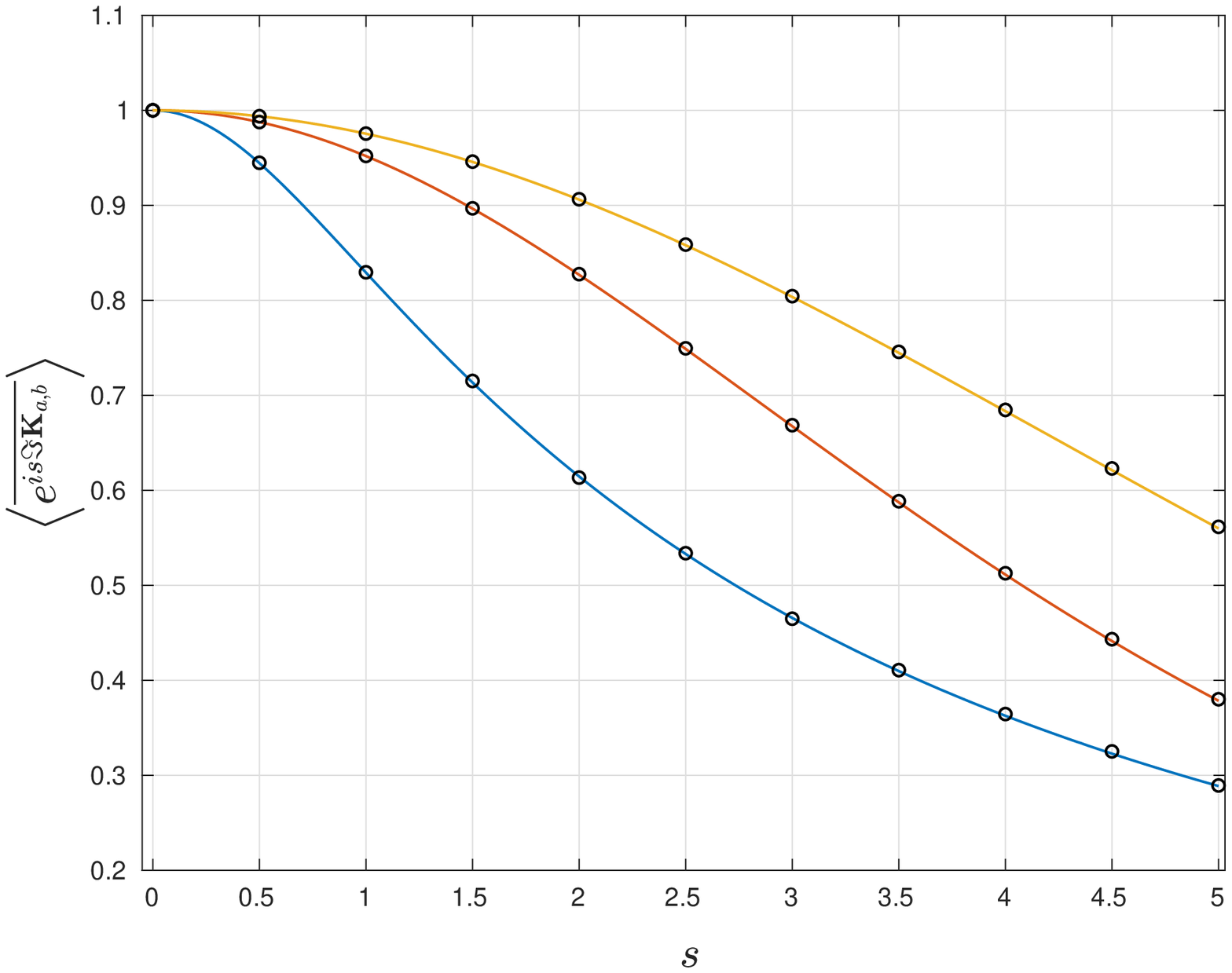}}
   {\includegraphics[width=.70\textwidth]{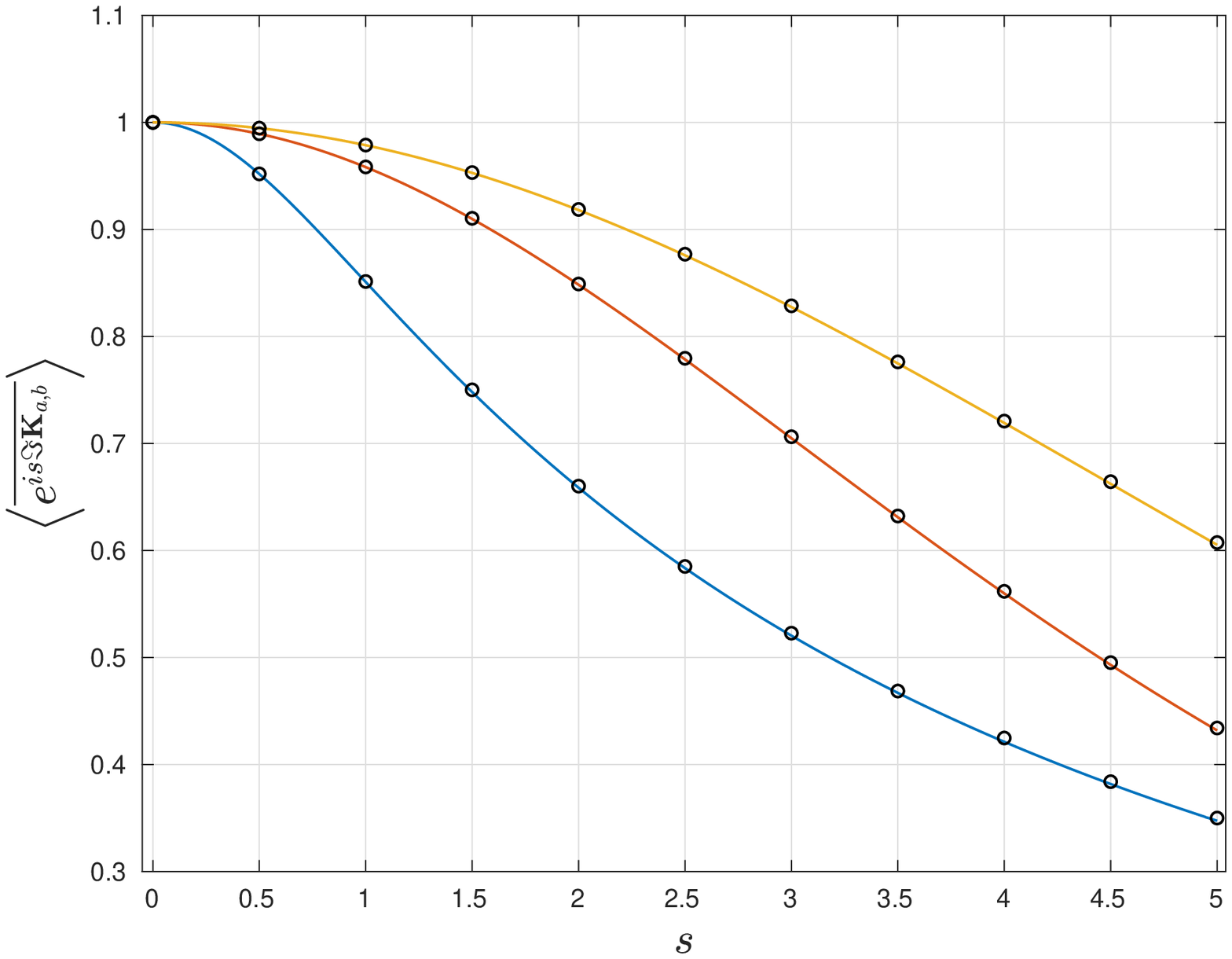}}\\
\label{pp10}
\caption{Characteristic function of $\Im K_{a,b}$ as given by (\ref{Im}) vs. numerical simulations for systems with preserved time-reversal invariance at different level of absorption: $\alpha=1(blue),5(red),10(yellow)$ (\# Samples 80000,N=100, circular markers) for $\lambda=0(a),1(b)$.}
\end{figure}

\begin{figure}[t]
\centering
 {\includegraphics[width=.70\textwidth]{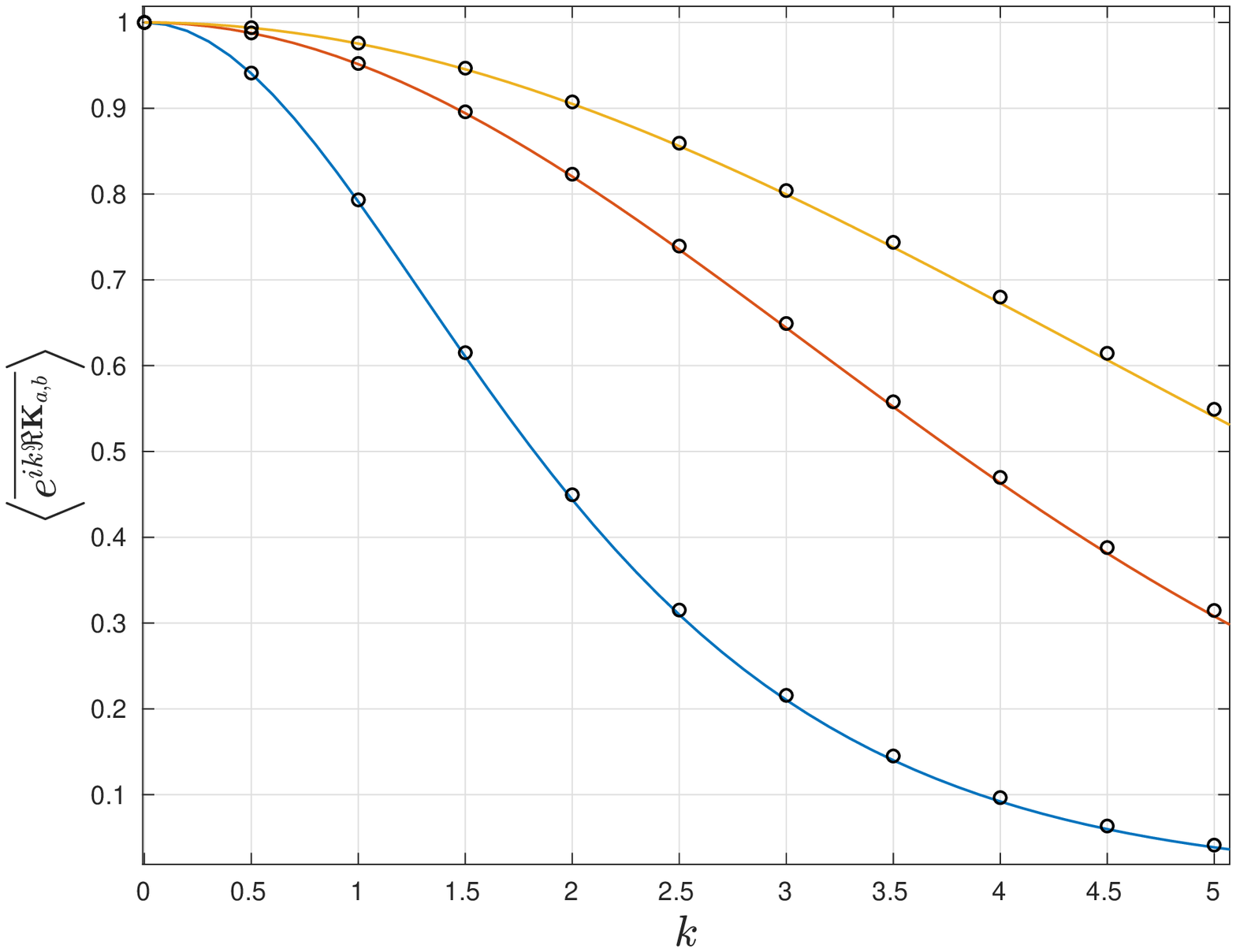}} \\
   {\includegraphics[width=.70\textwidth]{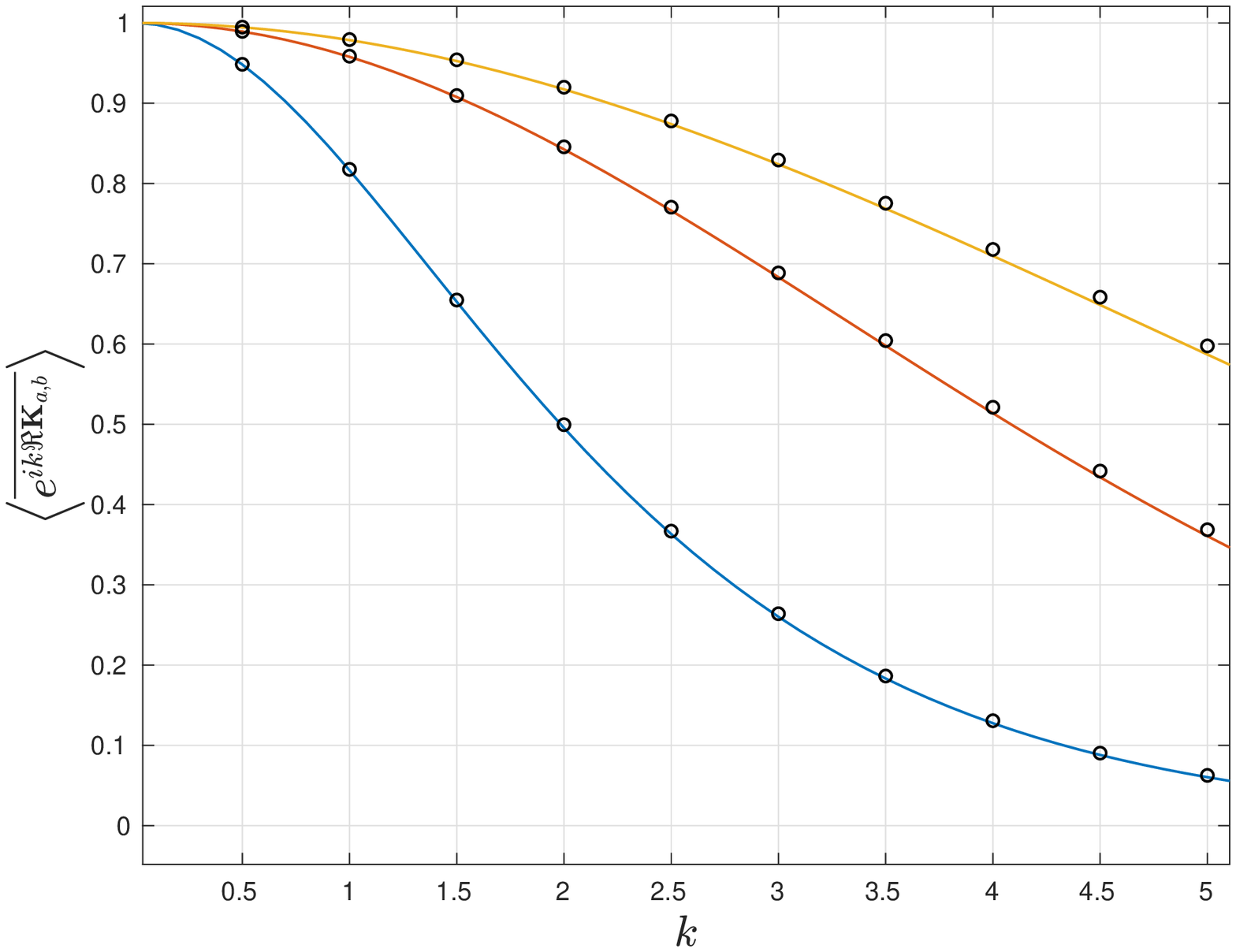}} \\
\label{pp10}
\caption{Characteristic function for $\Re K_{a,b}$ as given by (\ref{Re})  vs. numerical simulations for systems with preserved time-reversal invariance  at different level of absorption: $\alpha=1(blue),5(red),10(yellow)$ (\# Samples 50000, $N=300$, circular markers) for $\lambda=0(a),1(b)$.}
\end{figure}


\begin{figure}[!htb] \label{Gaularge}
  \centering
      \includegraphics[width=0.7\textwidth]{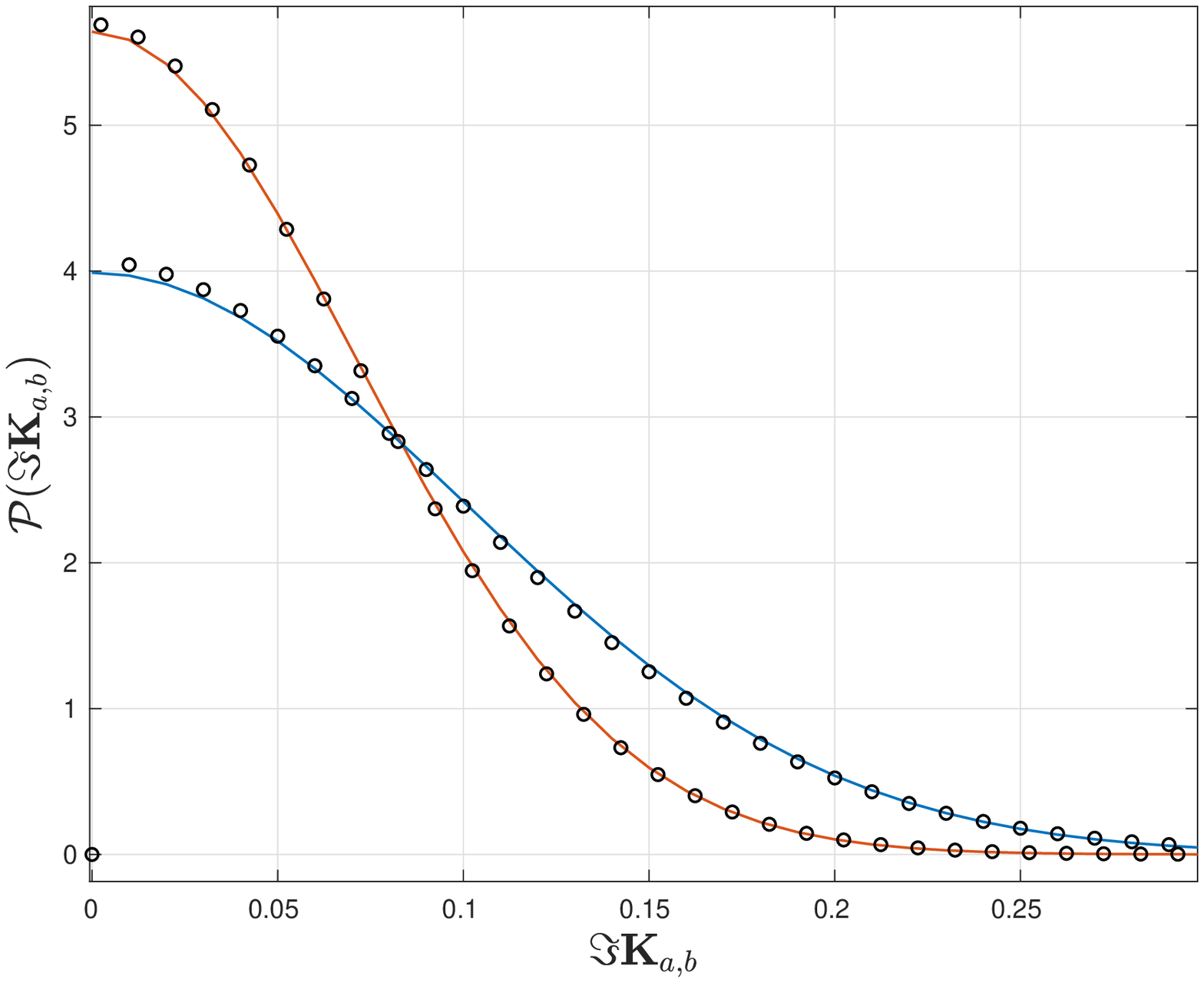}
             %

  \caption{Comparison between the probability density of $\Im K_{a,b}$ (Eq. (\ref{P11})) for large absorption $\alpha=50(blue),100(red)$, and the Gaussian distribution $\mathcal{N}(0,1/(2\alpha))$ (circular markers) ($\lambda=0$).}

\bigskip

      \includegraphics[width=0.7\textwidth]{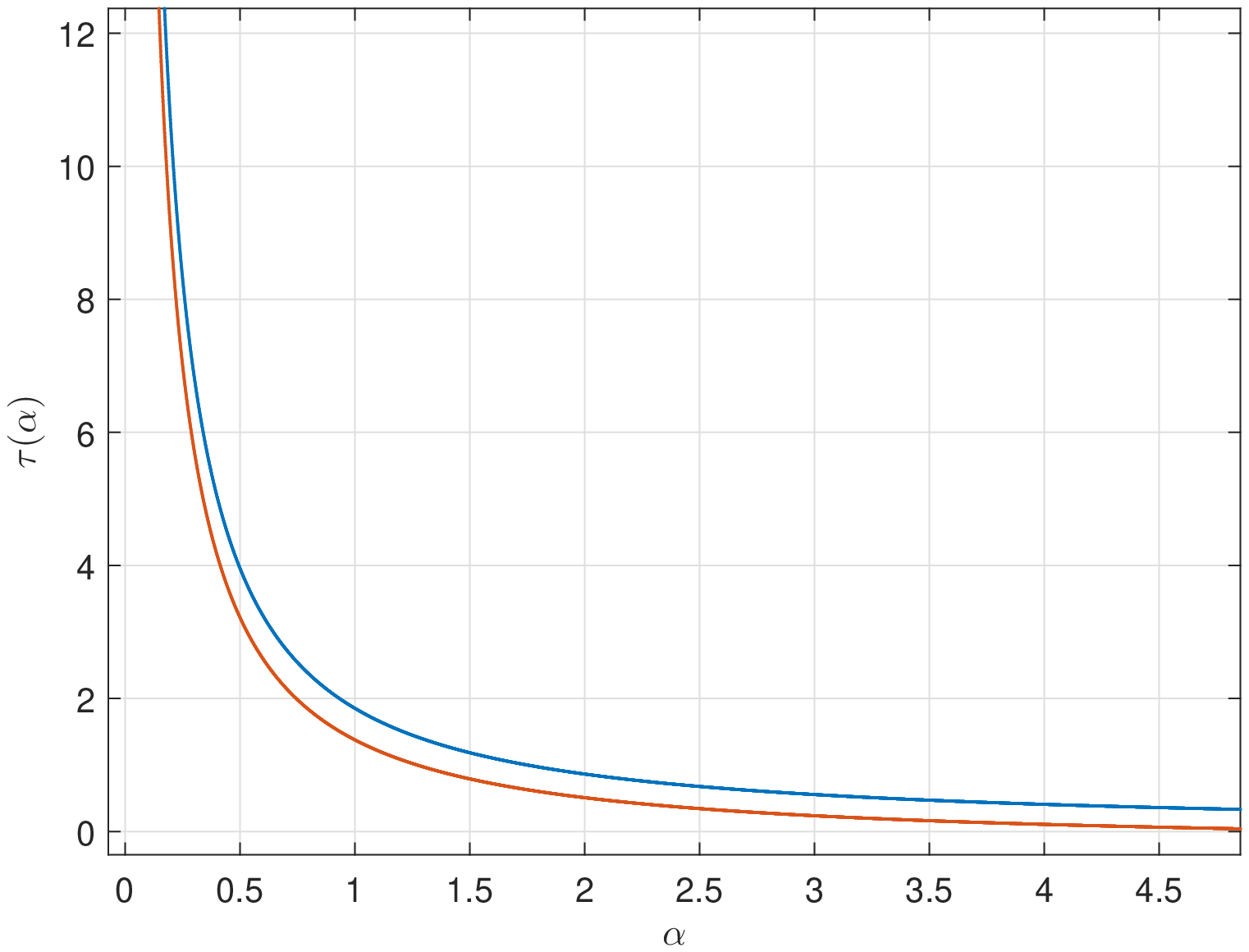}
  \caption{Behaviour of $\tau(\alpha)=\frac{\overline{\langle\left(\Im K_{a,b}\Re K_{a,b}\right)^2\rangle}}{\overline{\langle\left(\Im K_{a,b}\right)^2\rangle}\quad \overline{\langle\left(\Re K_{a,b}\right)^2\rangle}}-1, $ for $\lambda=0(blue),1(red)$}
\end{figure}

\section{Derivations of the main results}\label{A1c}
\subsection{Systems with broken time-reversal invariance} \label{A1cc}
Our starting point in this case is to write $\mathbf{K}_{a,b}$  in terms of the eigenvalues $\lambda_n$ and associated eigenvectors of the matrix $\mathbf{H}_N$:
$$
\mathbf{K}_{a,b}=\sum_{n=1}^N\frac{w_{a,n} w_{n,b}^{*}}{\lambda-\lambda_n+i\alpha/N}
$$
where $w_{n,a}=\left\langle{\bf w}_a|n\right\rangle, \, w_{n,b}^*=\left\langle n|{\bf w}_b\right\rangle$ are projection of the channel vectors on the eigenvectors $\left.|n\right\rangle, n=1,\ldots,N$.
Due to Gaussian nature of the channel vectors their projections on any system of orthonormal vectors are again Gaussian and independent.

We then aim to compute the following characteristic function
\[
\tilde{\mathcal{R}}(q,q^*)=
\Big\langle\overline{\exp{i\Big(\sum_{n=1}^Nq^{*}\frac{w_{a,n}w_{b,n}^*}{\lambda-\lambda_n+i\alpha/N}+q\frac{w_{a,n}^*w_{b,n}}
{\lambda-\lambda_n-i\alpha/N}\Big)}}\Big\rangle_{GUE(N)}
\]
 Evaluating the standard Gaussian integrals over $w_{a,n}$ and $w_{b,n}$ yields
\[
\tilde{\mathcal{R}}(q,q^*)=\left\langle \prod_{n=1}^N\frac{(\lambda-\lambda_n)^2+\alpha^2/N^2}{(\lambda-\lambda_n)^2+\alpha^2/N^2+|q|^2/N^2}\right\rangle_{GUE(N)}
\]
\[
=\Bigg\langle\frac{\det((\mathbf{H}_N-\lambda{\bf 1}_N)^2+\alpha^2/N^2{\bf 1}_N)}{\det((\mathbf{H}_N-\lambda{\bf 1}_N)^2+(|q|^2/N+\alpha^2/N^2){\bf 1}_N)}\Bigg\rangle_{GUE(N)}:=\tilde{\mathcal{R}}(|q|)
\]
When $N\to \infty$, the above object has been evaluated in \cite{FS03b}. Namely, defining the two-point kernel $\mathcal{S}$
via $$
\mathbb{S}(x-y)=\left\{\begin{array}{cc}
{\frac{e^{i\pi(x-y)}}{x-y}}, & \mbox{if } \Im{x}>0\\
{\frac{e^{-i\pi(x-y)}}{x-y}}, & \mbox{if } \Im{x}<0
  \end{array}\right.
$$
one than has the following representation:
$$
\tilde{\mathcal{R}}(|q|)=\frac{\rho^2(\lambda)|q|^4}{4\alpha\sqrt{\alpha^2+|q|^2}}\det[e^{-\phi(\lambda)(\xi_i-\eta_j)}
\mathbb{S}(\xi_i-\eta_j)]_{i,j=1,2},
$$
with $\xi_1=i\rho(\lambda)\sqrt{\alpha^2+|q|^2}$, $\xi_2=-i\rho(\lambda)\sqrt{\alpha^2+|q|^2}$, $\eta_1=i\rho(\lambda)\alpha$ and  $\eta_2=i\rho(\lambda)\alpha$.
 The probability density function for ${\bf K}_{ab}$ is then obtained by Fourier-transforming ($\mathcal{R}(|q|)=\tilde{\mathcal{R}}(|q/2|)$):
$$
\mathcal{P}(\mathbf{K}_{a,b},\mathbf{K}_{a,b}^*)=\int e^{-i(\Re{q}(\mathbf{K}_{a,b}+\mathbf{K}_{a,b}^*)/2+\Im{q}(\mathbf{K}_{a,b}-\mathbf{K}_{a,b}^*)/2)}\mathcal{R}(|q|)\frac{d\Re{q}\,d\Im{q}}{(2\pi)^2}
$$
Changing to polar coordinates, integrating out angular variables, performing obvious manipulations and finally rescaling leads eventually to:
\[
\mathcal{P}(\mathbf{K}_{a,b},\mathbf{K}_{a,b}^*)=\frac{\alpha^2}{\pi}\lim_{x\rightarrow2\alpha\pi\rho(\lambda)}\int_{0}^{\infty}dr\, \frac{r}{\sqrt{1+r^2}}J_0(2\alpha|\mathbf{K}_{a,b}|r)
\]
\[
\times \Big(\sinh(x)\Big(1+\frac{d^2}{dx^2}\Big)-2\cosh(x)\frac{d}{dx}\Big)\exp{(-x\sqrt{1+r^2})}.
\]
Changing $y=\sqrt{1+r^2}$ and using that $K_{1/2}(u)=\sqrt{\frac{\pi}{2u}}e^{-u}$, where $K_{\nu}(u)$ is the Bessel-Macdonald function of order $\nu$, allows to compute the integral, see \cite{GR}:
$$
\int_{0}^{+\infty}dr\frac{J_0(2\alpha|\mathbf{K}_{a,b}|r)r}{\sqrt{1+r^2}}\exp{(-x\sqrt{1+r^2})}
=\frac{e^{-\sqrt{x^2+4\alpha^2|\mathbf{K}_{a,b}|^2}}}{\sqrt{x^2+4\alpha^2|\mathbf{K}_{a,b}|^2}}.
$$


\subsection{Systems with broken time-reversal invariance and correlated channels.}
In this section we sketch the derivation in the case of channels correlated as described in the Sec {\it 2.1.2} and characterized via $2\times2$ complex matrix $\mathbf{C}^{-1}$. The characteristic function $\langle\exp{\frac{i}{2}(q\mathbf{K}_{ab}^{*}+q\mathbf{K}_{ab}^{*})}\rangle_{GUE(N)}$ is then given by the ensemble average of
$$
\prod_{n=1}^N\frac{N^2\det \mathbf{C}}{\pi^2}\int_{\mathbb{C}^2} \Big(\prod_{j=a,b}dw_{j,n}dw_{j,n}^*\Big)
\exp\Big\{
-N\left[\begin{array}{c}
w_{a,n} \\
w_{b,n} \\
\end{array}\right]^{\dag}\mathbf{C}
\left[\begin{array}{c}
w_{a,n} \\
w_{b,n} \\
\end{array}\right]+\frac{i}{2}\Big(q\frac{w_{a,n}^*w_{b,n}}{\delta_n^*}+q^*\frac{w_{a,n}w_{b,n}^*}{\delta_n}\Big)
\Big\}
$$
where we introduced the notation $\delta_n=\lambda-\lambda_n-i\alpha/N$. Performing the gaussian integrals over the channel variables allows to represent the characteristic function as
$$
\Big\langle\prod_{n=1}^N\frac{N^2\det\mathbf{C}}{N^2C_{11}C_{22}-(NC_{12}-i\frac{q}{2\delta_n^*})(NC_{12}^*-i\frac{q}{2\delta_n})}\Big\rangle_{GUE(N)}
$$
 which in turn can be equivalently represented as an average of the ratio of determinants over the GUE ensemble:
$$
\Big\langle\frac{\det((\lambda-i\alpha/N){\bf 1}_N-\mathbf{H})\det((\lambda+i\alpha/N){\bf 1}_N-\mathbf{H})}{\prod_{j=1,2}\det((\lambda{\bf 1}_N-\mathbf{H})+\frac{i}{2N}(\tilde{k}+(-1)^j\sqrt{\tilde{k}^2+4\tilde{s}}){\bf 1}_N)}\Big\rangle_{GUE(N)}
$$
where we denoted
$$
\tilde{k}=\frac{1}{2}\Big(\frac{C_{12}^*}{\det\mathbf{C}}q+\frac{C_{12}}{\det\mathbf{C}}q^*\Big)
$$
$$
\tilde{s}=\alpha^2+\frac{\alpha}{2}\Big(\frac{C_{12}^*}{\det\mathbf{C}}q-\frac{C_{12}}{\det\mathbf{C}}q^*\Big)+\frac{|q|^2}{4\det{\mathbf{C}}}
$$
Evaluating the ensemble average in the large$-N$ limit (see eq[4.9] in \cite{FS03a}) we then arrive at
the characteristic function given by:
$$
\mathcal{R}(q,q^*)=-\frac{(\varepsilon_1-\mu_1)(\varepsilon_1-\mu_2)(\varepsilon_2-\mu_1)(\varepsilon_2-\mu_2)}{(\varepsilon_1-\varepsilon_2)(\mu_1-\mu_2)}(N\rho(\lambda))^2\det\big[e^{-\frac{\lambda}{2\rho(\lambda)}(\xi_i-\eta_j)}\mathbb{S}(\xi_i-\eta_j)\big]_{i,j=1,2}
$$
where we used the notations
$$
\left\{\begin{array}{cc}
\varepsilon_j=\lambda+\frac{\xi_j}{N\rho(\lambda)}\\
\mu=\lambda+\frac{\eta}{N\rho(\lambda)} \\
\xi_{j}=\rho(\lambda)\Big((-1)^j\frac{1}{2}\Im\sqrt{\tilde{k}^2+4\tilde{s}}+\frac{i}{2}(\tilde{k}+(-1)^{j+1} \Re\sqrt{\tilde{k}^2+4\tilde{s}}\Big)\\
\eta_{j}=(-1)^j\alpha\rho(\lambda)i\\
  \end{array}\right.
$$
with $j=1,2$. Finally, after simple algebra we arrive at the final result given in Eq.(\ref{RFC}).

\subsection{Derivations for the case of preserved time-reversal invariance}\label{Ac22}

In full analogy with the previous section we  consider the characteristic function of $\mathbf{K}_{a,b}$:
$$
\mathcal{R}(q,q^{*})=\langle \overline{e^{i/2(q^{*}\mathbf{K}_{ab}+q\mathbf{K}_{ab}^{*})}}\rangle_{GOE(N)}
$$
with $q=k+i s\in\mathbb{C}$. The argument in the exponential can be written in the basis of the eigenvector of $\mathbf{H}$ as:
$$
q^{*}\mathbf{K}_{ab}+q\mathbf{K}_{ab}^{*}=\sum_{n=1}^{N}w_{a,n}w_{b,n}\Big(\frac{q^*(\lambda-\lambda_n-i\alpha/N)
+q(\lambda-\lambda_n+i\alpha/N)}{(\lambda-\lambda_n)^2+\alpha^2/N^2}\Big).
$$
 The integration over the real Gaussian variables $w_{a,n},w_{b,n}$ and simple rearranging leads to representation eq.(\ref{GOEeq}).

At the next step we set $\lambda=0$ in eq.(\ref{GOEeq}) and represent $\mathbf{H}_N$ as a block matrix:
\begin{equation}
\mathbf{H}_N=
  \left[\begin{array}{@{}cc|c@{}}
    H_{11} & H_{12} & \bm{h}_{1}^{T}\\
    H_{12} & H_{22} & \bm{h}_{2}^{T}\\\hline
   \bm{h}_1 & \bm{h}_2 &\mathbf{H}_{N-2}\\
  \end{array}\right],
  \label{matrix}
\end{equation}
where $\mathbf{H}_{N-2}$ is the $(N-2)\times(N-2)$ submatrix obtained by deleteing the first two columns and the first two rows of $\mathbf{H}_N$, while $\bm{h}_1$ and $\bm{h}_2$ are $(N-2)$ dimensional vectors. The numerator in eq.(\ref{GOEeq}) for $\lambda=0$ can be then written
using the Schur complement formula as
\[\det\left(\mathbf{H}_N^2+\frac{\alpha^2}{N^2}{\bf 1}_N\right)=\det\left(\mathbf{H}_{N-2}^2+\frac{\alpha^2}{N^2}{\bf 1}_N\right)|\Delta|^2
\]
 with:
\[
\Delta=\det\Bigg(
\left[\begin{array}{cc}
   H_{1,1}-i\frac{\alpha}{N} & H_{1,2} \\
   H_{1,2}  & H_{2,2}-i\frac{\alpha}{N}\\
\end{array}\right]-\left[
\begin{array}{c}
    \bm{h}_{1}^T \\
    \bm{h}_{2}^T \\
\end{array}\right]
\frac{1}{\mathbf{H}_{N-2}-i\frac{\alpha}{N}}
\left[\begin{array}{cc}
    \bm{h}_{1}&\bm{h}_{2}\\
\end{array}\right]\Bigg).
\]
The determinants in the denominator of eq. (\ref{GOEeq}) can be replaced by Gaussian integrals via $\int_{\mathbb{R}^N}d\bm{x}\exp\{-\bm{x}^T\mathbf{A}\bm{x}\}\propto1/\sqrt{\det\mathbf{A}}$ for $\Re A\succ 0$. As the result, the characteristic function for $\Im\mathbf{K}_{ab}$ at $\lambda=0$ can be represented by the following integral:
\begin{equation}\label{restore}
\fl \overline{\Big\langle e^{is\Im \mathbf{K}_{ab}}\Big\rangle}_{GOE(N)}\propto\Bigg\langle\int_{\mathbb{R}^{2N}}d\bm{x}_1d\bm{x}_2 \exp\Bigg(-\Tr\Bigg\{\mathbf{H}_N^2\mathbf{Q}+\sum_{j=1,2}\frac{\omega_j^2}{N^2}{\bf 1}_{N}\Big(\bm{x}_j\otimes\bm{x}_j^T\Big)
\Bigg\}\Bigg)
\end{equation}
\[
\times\det\Big(\mathbf{H}_{N-2}^2+\frac{\alpha^2}{N^2}{\bf 1}_N\Big)|\Delta|^2\Bigg\rangle_{GOE(N)},
\]
where here and afterwards we systematically suppress multiplicative constants and  restore them only in the end of the calculation. The symmetric $N\times N$ matrix $\bm{Q}=\bm{x}_1\otimes\bm{x}_1^T+\bm{x}_2\otimes\bm{x}_2^T$ satisfies $Rank(\bm{Q})=2$. Therefore its spectrum $\sigma(\bm{Q})$ is given generically by two positive eigenvalues $\{q_1>0,q_2>0\}$, with the rest $N-2$ eigenvalues being identically zero, implying ${\bf Q}={\bf O}\,\mbox{diag}(q_1,q_2,0,\ldots,0){\bf O}^T$ with ${\bf O}\in O(N)$. The integral over variables $\bm{x}_1$ and $\bm{x}_2$ has a particular invariant form and can be written in terms of a $2\times2$ symmetric matrix
\[
\tilde{\mathbf{Q}}=
\left[\begin{array}{cc}
|\bm{x}_1|^2 & \bm{x}_1\cdot\bm{x}_2 \\
 \bm{x}_1\cdot\bm{x}_2 & |\bm{x}_2|^2
\end{array}\right]
\]
 with $\sigma(\tilde{\mathbf{Q}})=\sigma(\mathbf{Q})\setminus\{0\}$.  Now one can apply the following
\\
\\
\textbf{Proposition } (see \cite{F2002a}): \textit{For real matrices $\mathbf{Q}$ of the form $\sum_{j=1}^{k}\bm{x}_j\otimes\bm{x}_j^T$, where ${\bf x}_j\in\mathbb{R}^{N}, \forall j$ the following identity holds:}
\begin{equation}
\fl \int_{\mathbb{R}^{kN}} \prod_{j=1}^{k}d\bm{x}_j\Theta(\sigma(\mathbf{Q}))=\frac{\pi^{-k(k-1)/4}}{\prod_{j=0}^{k-1}\Gamma(\frac{N-j}{2})}
\int_{\tilde{\mathbf{Q}}>0}d\tilde{\mathbf{Q}}(\det\tilde{\mathbf{Q}})^{(N-k-1)/2}\Theta(\sigma(\tilde{\mathbf{Q}})),
\end{equation}
\textit{where the integration in the right-hand side is over the space of the real symmetric positive definite matrices of dimension $k\times k$}.

\vspace{0.5cm}

In our case $k=2$ and the integral over $\tilde{\mathbf{Q}}$  can be rewritten introducing the spectrum $q_1,q_2$ as integration variables. This leads to:
\begin{equation}
\overline{\Big\langle e^{{is\Im \mathbf{K}_{a,b}}}\Big\rangle}_{GOE(N)}\propto\int_{0}^{\infty}\int_{0}^{\infty}dq_1\,dq_2(q_1q_2)^{\frac{N-3}{2}}|q_1-q_2|
\Phi(\sigma(\tilde{\mathbf{Q}});\alpha)
\label{intc}
\end{equation}
\[
\times\int_{O(2)} d\mu(\mathbf{O})\exp{\Big(-\frac{1}{N^2}\Tr\left[\begin{array}{cc}
\omega_1^2 &0 \\
0 &\omega_2^2
\end{array}\right]\mathbf{O}\left[\begin{array}{cc}
q_1 &0 \\
0 &q_2
\end{array}\right]\mathbf{O}^{T}\Big)},
\]
where we denoted
\begin{equation}
\Phi(\sigma(\tilde{\mathbf{Q}}),\alpha)=\langle\det(\mathbf{H}^2_N+\alpha^2/N^2)\exp{(-\Tr\mathbf{H}^2_N\mathbf{Q})}\rangle_{GOE(N)}.
\label{avg}
\end{equation}
Parametrizing $\mathbf{O}\in O(2)$  as $\left[\begin{array}{cc} \cos{\phi} &\sin{\phi} \\-\sin{\phi} &\cos{\phi}\end{array}\right]$ the group integration in eq.(\ref{intc}) is equivalent to:
\begin{equation}\label{o2int}
\int_{0}^{\pi/2} d\phi\, e^{\cos^2{\phi}(\omega_1^2q_1+\omega_2^2q_2)+\sin^2{\phi}(\omega_1^2q_2+\omega_2^2q_1)}
=\frac{\pi}{2}\,e^{-\frac{\alpha^2}{N^2}(q_1+q_2)}J_{0}\Big(\frac{\alpha S}{N^2}(q_1-q_2)\Big),
\end{equation}
where $J_0(x)$ is the Bessel function of the first kind of order $0$.

The expectation over GOE ensemble in eq.(\ref{avg}) can be now evaluated directly by employing the block structure presented in eq.(\ref{matrix}).
Introducing the notations $\mathbf{M}=(\mathbf{H}_{N-2}-i\frac{\alpha}{N})^{-1}$ and $\mathbf{M}^{*}=(\mathbf{H}_{N-2}+i\frac{\alpha}{N})^{-1}$ one finds  that $\Phi$ can be written in the following form
$$
\Phi(q_1,q_2;\alpha)=\sum_{m,n,p}u_{m,n,p}(q_1,q_2,\alpha)\Big\langle\Big(Tr(\mathbf{M})^{m}(\mathbf{M}^{*})^{n}\Big)^p\Big\rangle_{GOE(N-2)}
$$
for some coefficients $u_{m,n,p}(q_1,q_2,\alpha)$ with $0\le m,n\le2,0\le p\le 4$.

 To see this we proceed as follows. We notice that $\Tr\left(\bm{H}^2_{N}(\bm{Q}+\frac{N}{4}{\bf 1}_{N})\right)$ is a quadratic polynomial in the entries of $\bm{H}_{N}$:
$$
\Tr\left(\bm{H}^2_{N}(\bm{Q}+\frac{N}{4}{\bf 1}_{N})\right)=\alpha_{1}H_{11}^2+\alpha_{2}H_{22}^2+\alpha_{12}H_{12}^2+\beta_1|\bm{h}_1|^2+\beta_2|\bm{h}_2|^2+\frac{N}{4}\Tr
\left(\bm{H}_{N-2}^2\right)
$$
with
\begin{equation}\label{alphabeta}
\fl \alpha_{1}=q_1+N/4,\quad \alpha_{2}=q_2+N/4,\quad \alpha_{12}=\alpha_{1}+\alpha_{2},\quad \beta_1=q_1+N/2,\quad \beta_2=q_2+N/2.
\end{equation}
Further we have
\[
\Delta\Delta^*=|H_{11}H_{22}-Z_{11}H_{22}-H_{11}Z_{22}+Z_{11}Z_{22}-H_{12}^2-Z_{12}^2+2H_{12}Z_{12}|^2,
\] where $Z_{ij}=i\alpha/N\delta_{ij}+\Tr(\mathbf{M}(\mathbf{h}_j\otimes\mathbf{h}_i^T))$. The Gaussian integrals over  $H_{1,1},H_{1,2},H_{2,2}$ can be performed using the identity
\begin{equation}
\fl \int_{-\infty}^{+\infty}e^{-ax^2}|cx^2+bx+d|^2dx=\frac{\sqrt{\pi}}{4a^{5/2}}(3|c|^2+2a|b|^2+2a(dc^{*}+d^{*}c)+4a^2|d|^2), \quad a>0, \, \mbox{and}\,\, b,c,d\in\mathbb{C}.
\label{iint}
\end{equation}
In this way we start with integrating out $H_{11}$ via
\[
\fl \int e^{-\alpha_1H_{11}^2}|\Delta|^2=\frac{\sqrt{\pi}}{2\alpha_1^{3/2}}
\left(|H_{22}-Z_{22}|^2+2\alpha_1|-Z_{11}H_{22}+Z_{11}Z_{22}-H_{12}^2-Z_{12}^2+2H_{12}Z_{12}|^2\right)
\]
and then similarly integrate over $H_{22}$ and $H_{12}$. This finally yields
\[
\Phi(q_1,q_2;\alpha)\propto\int \,d\mathbf{h}_1d\mathbf{h}_2d\mathbf{H}_{N-2}\, e^{-\beta_1|\mathbf{h}_1|^2-\beta_2|\mathbf{h}_2|^2-\frac{N}{4}
\Tr\mathbf{H}^2_{N-2}}\det\left(\mathbf{H}_{N-2}^2+\frac{\alpha^2}{N^2}{\bf 1}_{N-2}\right)
\]
\begin{equation}\label{lala}
\fl \times\Big(a_1+a_2|Z_{11}|^2+a_3|Z_{22}|^2+a_4|Z_{12}|^2+2a_5\Re(Z_{11}Z_{22}-Z_{12}^2)+a_6|Z_{11}Z_{22}-Z_{12}^2|^2\Big),
\end{equation}
where
\[
a_1=\frac{1}{\alpha_{12}^{1/2}}+3\frac{\alpha_1\alpha_2}{\alpha_{12}^{5/2}}, \, a_2=\frac{2\alpha_1}{\alpha_{12}^{1/2}} ,\, \, a_3=\frac{2\alpha_2}{\alpha_{12}^{1/2}}
\]
\[
a_4=\frac{8\alpha_1\alpha_2}{\alpha_{12}^{5/2}}, \, \, a_5=-2\frac{\alpha_1\alpha_2}{\alpha_{3/2}}, \, \, a_6=4\frac{\alpha_1\alpha_2}{\alpha_{12}^{1/2}}
\]
Remembering the definition of $Z_{ij}$, the integration in eq (\ref{lala}) over $\bm{h}_1$ and $\bm{h}_2$ relies on the following identities valid for $\beta_1,\beta_2>0$:
\[
\int_{\mathbb{R}^{N-2}}(\bm{h}_1^TM\bm{h}_1)e^{-\beta_1\bm{h}_1^2}d\bm{h}_1=\Big(\frac{\pi}{\beta_1}\Big)^{(N-2)/2}\frac{1}{2\beta_1}\Tr\mathbf{M}
\]
and
\[
\fl \int_{\mathbb{R}^{N-2}}e^{-\beta_1\bm{h}_1^2}\,\left(\bm{h}_1^T\mathbf{M}_1\bm{h}_1\right)\left(\bm{h}_1^T\mathbf{M}_2\bm{h}_1\right)
d \bm{h}_1=\frac{1}{4}\frac{\pi^{(N-2)/2}}{\beta_1^{(N-2)/2}}\frac{1}{\beta_1^2}(\Tr \mathbf{M}_1\Tr \mathbf{M}_2+2\Tr \mathbf{M}_1\mathbf{M}_2),
\]
as well as
\[
\fl \int_{\mathbb{R}^{2(N-2)}}e^{-\beta_1\bm{h}_1^2-\beta_2 \bm{h}_2^2}\,\left(\bm{h}_1^T\mathbf{M}_1\bm{h}_2\right)\left(\bm{h}_1^T\mathbf{M}_2\bm{h}_2\right)\, d\bm{h}_1d\bm{h}_2=\Big(\frac{\pi^2}{\beta_1\beta_2}\Big)^{(N-2)/2}\frac{1}{4\beta_1\beta_2}\Tr \mathbf{M}_1\mathbf{M}_2,
\]
and
 \[
\int_{\mathbb{R}^{2(N-2)}}e^{-\beta_1\bm{h}_1^2-\beta_2 \bm{h}_2^2}\left(\bm{h}_1^T\mathbf{M}_1\bm{h}_2\right)^2\left(\bm{h}_1^T\mathbf{M}_2\bm{h}_2\right)^2\,d\bm{h}_1d\bm{h}_2
\]
\[
\fl =\frac{1}{16}\Big(\frac{\pi^2}{\beta_1\beta_2}\Big)^{(N-2)/2}\frac{1}{\beta_1^2\beta_2^2}(\Tr \mathbf{M}_1^2\Tr \mathbf{M}_2^{2}+4\Tr \mathbf{M}_1^2\mathbf{M}_2^{2}+2\Tr^2 \mathbf{M}_1\mathbf{M}_2+2\Tr(\mathbf{M}_1\mathbf{M}_2)^2),
\]
and finally
\[
\int_{\mathbb{R}^{2(N-2)}}e^{-\beta_1\bm{h}_1^2-\beta_2 \bm{h}_2^2}\left(\bm{h}_1^T\mathbf{M}\bm{h}_2\right)^2\left(\bm{h}_1^T\mathbf{M}^{*}\bm{h}_1\right)\,d\bm{h}_1d\bm{h}_2
\]
\[
=\frac{1}{8}\Big(\frac{\pi^2}{\beta_1\beta_2}\Big)^{(N-2)/2}\frac{1}{\beta_1^2\beta_2}(\Tr \mathbf{M}^2\Tr \mathbf{M}^{*}+2\Tr(\mathbf{M}^2\mathbf{M}^{*}))
\]
and
\[
\int_{\mathbb{R}^{2(N-2)}}e^{-\beta_1\bm{h}_1^2-\beta_2\bm{h}_2^2}\left(\bm{h}_1^T\mathbf{M}\bm{h}_2\right)^2
\left(\bm{h}_1^T\mathbf{M}^{*}\bm{h}_1\right)\left(\bm{h}_2^T\mathbf{M}^{*}\bm{h}_2\right)
\, d\bm{h}_1d\bm{h}_2
\]
\[
=\frac{1}{16}\Big(\frac{\pi^2}{\beta_1\beta_2}\Big)^{(N-2)/2}\frac{1}{\beta_1^2\beta_2^2}(\Tr \mathbf{M}^2\Tr^2\mathbf{M}^{*}+4\Tr \mathbf{M}^2\mathbf{M}^{*}\Tr \mathbf{M}^{*}+4\Tr (\mathbf{M}\mathbf{M}^{*})^2).
\]

Performing in this way integrations over $\bm{h}_1$ and $\bm{h}_2$ leads to the following cumbersome intermediate expression:
\[
\Phi(q_1,q_2;\alpha)\propto\Big(\frac{\pi^2}{\beta_1\beta_2}\Big)^{(N-2)/2}\int d\mathbf{H}_{N-2}e^{-\frac{N}{4J^2}\Tr\mathbf{H}^2_{N-2}}\det(\mathbf{H}_{N-2}^2+\frac{\alpha^2}{N^2}{\bf 1}_{N-2})
\]
\[
\times\left\{u_1+u_2\Tr(\mathbf{M}-\mathbf{M}^*)+u_3\Tr\mathbf{M}\mathbf{M}^*+2u_4\Re(\Tr^2\mathbf{M}-\Tr\mathbf{M}^2)+u_5\Tr\mathbf{M}
\Tr\mathbf{M}^*\right.
\]
\[
+u_6\Big(\Tr(\mathbf{M}-\mathbf{M}^*)(\Tr\mathbf{M}\Tr\mathbf{M}^*+2\Tr\mathbf{M}\mathbf{M}^*)
\]
\[
+\Tr\mathbf{M}^2\Tr\mathbf{M}^*+2\Tr\mathbf{M}^2\mathbf{M}^{*}-(\Tr\mathbf{M}^{*2}\Tr\mathbf{M}+2\Tr\mathbf{M}^{*2}\mathbf{M})\Big)
\]
\[
+u_7\Big((\Tr\mathbf{M}\Tr\mathbf{M}^*+2\Tr\mathbf{M}\mathbf{M}^*)^2+\Tr\mathbf{M}^2\Tr\mathbf{M}^{*2}+6\Tr(\mathbf{M}\mathbf{M})^2
\]
\[
\left. +2(\Tr\mathbf{M}\mathbf{M}^*)^2-2\Re(\Tr\mathbf{M}^2(\Tr\mathbf{M}^*)^2+4\Tr\mathbf{M}^2\mathbf{M}^{*}
\Tr\mathbf{M}^{*}+4\Tr(\mathbf{M}\mathbf{M}^{*})^2)\Big)\right\},
\]
with:
$$
u_1=a_1+\frac{\alpha^2}{N^2}a_2+\frac{\alpha^2}{N^2}a_3-2\frac{\alpha^2}{N^2}a_5+\frac{\alpha^4}{N^4}a_6,
$$
$$
u_2=-i\frac{\alpha}{N}\frac{a_2}{2\beta_1}-i\frac{\alpha}{N}\frac{a_3}{2\beta_2}+i\frac{\alpha a_5}{(2N)}\big(\frac{1}{\beta_1}+\frac{1}{\beta_2}\big)-i\frac{\alpha^3 a_6}{(2N^3)}\big(\frac{1}{\beta_1}+\frac{1}{\beta_2}\big),
$$
$$
u_3=\frac{a_2}{2\beta_1^2}+\frac{a_3}{2\beta_2^2}+\frac{a_4}{4\beta_1\beta_2}+\frac{a_6\alpha^2}{2N^2}
\big(\frac{1}{\beta_1^2}+\frac{1}{\beta_2^2}\big),
$$
$$
u_4=\frac{a_5}{4\beta_1\beta_2}-\frac{\alpha^2}{4N^2}\frac{a_6}{\beta_1\beta_2}\mbox{;}\, u_5=\frac{a_2}{4\beta_1^2}+\frac{a_3}{4\beta_2^2}+\frac{\alpha^2 a_6}{4N^2}\big(\frac{1}{\beta_1^2}+\frac{1}{\beta_2^2}\big)+\frac{\alpha^2}{2N^2}\frac{a_6}{\beta_1\beta_2},
$$
$$
u_6=\frac{i\alpha}{8N}a_6\big(\frac{1}{\beta_1^2\beta_2}+\frac{1}{\beta_1\beta_2^2}\big)\mbox{;}\quad u_7=\frac{a_6}{16\beta_1^2\beta_2^2}.
$$
In this way evaluating  $\Phi(\sigma(\tilde{\mathbf{Q}}),\alpha)$ is reduced to performing  $GOE(N-2)$ averaging of polynomials of traces for $\mathbf{M}=\left(\mathbf{H}_{N-2}-i\frac{\alpha}{N}\right)^{-1}$ and its complex conjugate. Observing that $\mathbf{M}-\mathbf{M}^{*}=2i\frac{\alpha}{N}\mathbf{M}\mathbf{M}^{*}$, each monomial can be rewritten as a combination of derivatives of characteristic polynomials, i.e. $\partial_{\xi}^m\det(\mathbf{H}_{N-2}-(\xi\pm i\alpha/N))|_{\xi=0}$ for some $m>0$. Exploiting this the final integration over $\mathbf{H}_{N-2}$ is performed as follows. First we introduce the correlation function of the product of two characteristic polynomials, which in the limit $N\gg 1$ takes the form, see e.g. \cite{Kosters2008}:
$$
\Big\langle\det\Big(\mathbf{H}_{N-2}-i\frac{\alpha}{N}-\xi_{+}\Big)\det\Big(\mathbf{H}_{N-2}+i\frac{\alpha}{N}-\xi_{-}\Big)
\Big\rangle_{GOE(N-2)}
$$
\begin{equation}
\propto \frac{-f(\xi_{+}-\xi_{-})\cos\Big(f(\xi_{+}-\xi_{-})\Big)+\sin\Big(f(\xi_{+}-\xi_{-})\Big)}{f^3(\xi_{+}-\xi_{-})}
:=C_{SP}(\xi_{+}-\xi_{-})
\label{c}
\end{equation}
with $f(\xi)=2i\alpha+N\xi$. Now by employing the identities

\[
\fl \frac{d}{d\xi}\det(\mathbf{H}_{N-2}-(\xi\pm i\alpha/N){\bf 1}_N)=-\Tr\Big(\mathbf{H}_{N-2}-(\xi\pm i\alpha/N){\bf 1}_N\Big)^{-1}\det(\mathbf{H}_{N-2}-(\xi\pm i\alpha/N){\bf 1}_N)
\]
and
\[
\frac{d}{d\xi}\Tr\Big(\mathbf{H}_{N-2}-(\xi\pm i\alpha/N){\bf 1}_N\Big)^{-k}=k\Tr\Big(\mathbf{H}_{N-2}-(\xi\pm i\alpha/N){\bf 1}_N\Big)^{-(k+1)}
\]

 one is able to represent $\Phi$ in the following form:
\[
\Phi(q_1,q_2;\alpha)\propto\lim_{\delta\rightarrow0}\sum_{j=0}^{4}b_j(q_1,q_2,\alpha)\,\hat{D}_j C_{SP}(\delta)
\]
where the coefficients $b_j$ are given by the following expressions:
\[
b_0=\frac{1}{\sqrt{\alpha_{12}}}+3\frac{\alpha_1\alpha_2}{\alpha_{12}^{5/2}}+2\frac{\alpha^2}{N^2}
\frac{\alpha_1+\alpha_2}{\sqrt{\alpha_{12}}}+4\frac{\alpha^2}{N^2}\frac{\alpha_1\alpha_2}{\alpha_{12}^{3/2}}
+4\frac{\alpha^4}{N^4}\frac{\alpha_1\alpha_2}{\sqrt{\alpha_{12}}},
\]
\[
b_1=-\frac{i}{\sqrt{\alpha_{12}}}\frac{\alpha}{N}\Big(\frac{\alpha_1}{\beta_1}+\frac{\alpha_2}{\beta_2}\Big)-i\frac{\alpha}{N}\frac{\alpha_1\alpha_2}{\alpha_{12}^{3/2}}\Big(\frac{1}{\beta_1}+\frac{1}{\beta_2}\Big)-i2\frac{\alpha^3}{N^3}\frac{\alpha_1\alpha_2}{\sqrt{\alpha_{12}}}\Big(\frac{1}{\beta_1}+\frac{1}{\beta_2}\Big)+
\]
\[
-\frac{i N}{2\alpha}\Big(\frac{1}{\sqrt{\alpha_{12}}}\Big( \frac{\alpha_1}{\beta_1^2}+\frac{\alpha_2}{\beta_2^2}\Big)+2\frac{\alpha_1\alpha_2}{\alpha_{12}^{3/2}\beta_1\beta_2}
+2\frac{\alpha^2}{N^2}\frac{\alpha_1\alpha_2}{\sqrt{\alpha_{12}}}\Big(\frac{1}{\beta_1^2}+\frac{1}{\beta_2^2}\Big)\Big),
\]
\[
\fl b_2=-\frac{\alpha_1\alpha_2}{\alpha_{12}^{3/2}\beta_1\beta_2}-2\frac{\alpha^2}{N^2}\frac{\alpha_1\alpha_2}{\sqrt{\alpha_{12}}\beta_1\beta_2}-\Big(\frac{\alpha_{1}}{2\sqrt{\alpha_{12}}\beta_1^2}+\frac{\alpha_{2}}{2\sqrt{\alpha_{12}}\beta_2^2}+\frac{\alpha^2}{N^2}\frac{\alpha_1\alpha_2}{\sqrt{\alpha_{12}}}\Big(\frac{1}{\beta_1^2}+\frac{1}{\beta_2^2}+\frac{2}{\beta_1\beta_2}\Big)\Big)
\]
and finally
\[
b_3=i\frac{\alpha}{2N}\frac{\alpha_1\alpha_2}{\sqrt{\alpha_{12}}}\Big(\frac{1}{\beta_1\beta_{2}^2} +\frac{1}{\beta_2\beta_{1}^2}\Big)\quad \mbox{ and } \quad b_4=\frac{\alpha_1\alpha_2}{4\sqrt{\alpha_{12}}\beta_1^2\beta_2^2},
\]
whereas $\hat{D}_j$ stand for the following differential operators:
\[
\hat{\mathcal{D}}_0=1\mbox{, } \quad \hat{\mathcal{D}}_1=-2\partial_{\delta}\mbox{, }\quad \hat{\mathcal{D}}_2=\partial^2_{\delta}\mbox{, }\quad \hat{\mathcal{D}}_3=-2\partial^3_{\delta}+\frac{4Ni}{\alpha}\partial^2_{\delta}-\frac{2N^2}{\alpha^2}\partial_{\delta}
\]
and
\[
\hat{\mathcal{D}}_4=\partial^4_{\delta}-\frac{N^2}{\alpha^2}\Big(2\partial^2_{\delta}
+\frac{iN}{\alpha}\partial_{\delta}-\frac{4Ni}{\alpha}\partial^3_{\delta}\Big)
\]
where $\partial^{k}_{\delta}:=\frac{\partial^k}{\partial \delta^k}$. Now by
rescaling $q_{1,2}\rightarrow N^2q_{1,2}$ and recalling the definitions (\ref{alphabeta}) we see that for $N\gg 1$ to the leading order
we can replace
\[
(\beta_1\beta_2)^{N/2}\beta_1\beta_2\approx e^{-1/4(q^{-1}_1+q^{-1}_2)}(q_1q_2)^{1-N/2},
\]
as well as  $\alpha_1\approx N^2 q_1$ and $\alpha_2\approx N^2q_2$.

  Finally, we can restore the suppressed proportionality constant in equation (\ref{restore}) by noticing that for any $N$ $\overline{\langle\exp{\left(is\Im \mathbf{K}_{a,b}\right)}\rangle}_{GOE(N)}=1$ for either $\alpha=0$ or $s=0$. Note however that in the eq.(\ref{intc}) the limit $\alpha\rightarrow 0$ and the integration over $q_{1,2}$ do not commute. Therefore the constant of proportionality must be a function of $\alpha$. The only unknown factor comes from eq.(\ref{c}). To this end we consider $N$ to be odd for simplicity and further consider the limit $\xi_{\pm}\rightarrow0$ and $\alpha\rightarrow0$.  We then have, see
  \cite{Del2000}:
  \begin{equation}\label{ScaleDet}
\fl \langle|\det\mathbf{H}_{N-2}|^{r-1}\rangle_{GOE(N-2)}=N^{\frac{(N-2)(r-1)}{2}}\,2^{r-1}\frac{\Gamma(r/2)}{\Gamma(1/2)}\prod_{j=1}^{(N-3)/2}
2^{r-1}\frac{\Gamma(r+j-1/2)}{\Gamma(j+1/2)}
\end{equation}
with $r>1$. This allows to restore the multiplicative constant and obtain the characteristic function as in eq.(\ref{Im}). Further inverting the Fourier transform we obtain the probability density function of $\Im \mathbf{K}_{a,b}$. In doing this it is useful to notice that:
\[
\mathcal{F}^{-1}\Big[J_{0}\Big(\alpha s(q_1-q_2)\Big)\Big]:=\frac{1}{2\pi}\int_{-\infty}^{+\infty}ds e^{{-i s \Im \mathbf{K}_{ab}}}J_{0}\Big(\alpha s(q_1-q_2)\Big)=
\]
\[
\frac{1}{\pi}\frac{{\bf 1}(\Omega)}{\sqrt{\alpha^2(q_1-q_2)^2-\Im^2 \mathbf{K}_{a,b}}},
\]
where ${\bf 1}(\Omega)$ is the indicator function of the set $\Omega=\{(q_1,q_2)\in\mathbb{R}_{+}^2|\alpha^2(q_1-q_2)^2>\Im^2 K_{a,b}\}$.
The form of the set $\Omega$ suggests to introduce a new rescaled variable $u=\alpha^{-1}\Im K_{a,b}$ to arrive to eq.(\ref{P11}) .

\subsection{Derivation of eq.(\ref{Re})}
The derivation of the characteristic function for $\Re \mathbf{K}_{a,b}$ follows very similar lines and we only briefly sketch it here. Assuming $\lambda=0$, it is sufficient to observe that
\begin{equation}\label{redis}
\fl \overline{\Big\langle e^{ik\Re \mathbf{K}_{a,b}}\Big\rangle}_{GOE(N)} =\Bigg\langle\frac{\det\left(\mathbf{H}_N^2+\frac{\alpha^2}{N^2}{\bf 1}_N\right)}
{\Pi_{l=1,2}\det^{1/2}\left(\mathbf{H}_N^2+(-1)^l\,i\frac{k}{N}{\bf H}_N+\frac{\alpha^2}{N^2}{\bf 1}_N\right)}\Bigg\rangle_{GOE(N)}\,.
\end{equation}
Noticing that the denominator can be rewritten as:
$$
\det(\mathbf{H}_N^2+\tilde{\omega}_1^2)^{-1/2}\det(\mathbf{H}_N^2+\tilde{\omega}_2^2)^{-1/2}
$$
with $\tilde{\omega}_1^2=(k/2+\sqrt{\alpha^2+k^2/4})^2/N^2$ and  $\tilde{\omega}_1^2=(k/2-\sqrt{\alpha^2+k^2/4})^2/N^2$ we conclude that the GOE expectation in (\ref{redis}) can be obtained from the corresponding expression  eq.(\ref{restore}) for the characteristic function of $\Im \mathbf{K}_{a,b}$ by simply replacing $\omega_{1,2}$ in eq.(\ref{restore}) with the new values $\tilde{\omega}_{1,2}$ as above. The only difference comes from the integration over the orthogonal group of $2\times 2$ matrices in eq.(\ref{o2int}) which is replaced with
$$
I=\frac{\pi}{2}\exp{\Big(-\frac{1}{2N^2}(k^2+2\alpha^2)(q_1+q_2)\Big)}J_0\Big(i\frac{k}{N^2}\sqrt{\alpha^2+\frac{k^2}{4}}(q_1-q_2)\Big).
$$
Note that we found challenging to obtain an explicit probability density of $\Re\mathbf{K}_{a,b}$ by inverting the Fourier transform as the integral
$$
\frac{1}{2\pi}\int_{-\infty}^{\infty}\, e^{{-i k \Re \mathbf{K}_{ab}}}\exp{\Big(-\frac{(q_1+q_2)}{2N^2}k^2\Big)}I_0\Big(\frac{k}{N^2}\sqrt{\alpha^2+\frac{k^2}{4}}(q_1-q_2)\Big)\,dk
$$
does not seem to have a simple closed form expression, and can be evaluated only numerically.

\section{Appendices}

\subsection{Appendix A}\label{JR2}
 In this appendix we quote an explicit integral representation from \cite{FN2015}:
$$
\fl \overline{\Big\langle e^{is\Im \mathbf{K}_{a,b}}\Big\rangle}_{GOE(N)}\propto \int_{\mathbf{Q}\succeq0} d\mathbf{Q}(\det\mathbf{Q})^{(N-5)/2}\exp\Big(-\frac{N}{4J^2}\Tr(\mathbf{Q}\mathbf{L})^2+i\frac{N}{2J^2}
\Tr(\mathbf{Q}\mathbf{L}\mathbf{M})\Big)
$$
$$
\times \, \int_{\mathbb{R}} dr_1\int_{\mathbb{R}} dr_2\exp\Big(-\frac{N}{2J^2}(r_1^2+r_2^2-i2\lambda(r_1+r_2))\Big)\frac{(r_1r_2)^{N-4}}{(2i\alpha)^3}
$$
\begin{equation}
\times \, \prod_{j=1}^4(r_1+\lambda_{j})(r_2+\lambda_{j})\exp\Bigg(\frac{N}{J^2}(\lambda^2-\alpha^2/N^2)\Bigg)
\label{SSym}
\end{equation}
$$
\times\, \Bigg(\frac{2i\alpha}{N(r_1-r_2)}\cos\Big(\frac{2i\alpha(r_1-r_2)}{2J^2}\Big)-\frac{2J^2}{N}
\sin\Big(\frac{2i\alpha(r_1-r_2)}{2J^2}\Big)\Bigg).
$$
In the above $\mathbf{Q}$ is $4\times4$ positive definite real symmetric matrix, $\mathbf{L}=\mbox{diag}(+1,+1,-1,-1)$ and $\mathbf{M}=\mbox{diag}(\lambda+\frac{i}{N}\sqrt{\alpha^2+i\alpha s},\lambda+\frac{i}{N}\sqrt{\alpha^2-i\alpha s},\lambda-\frac{i}{N}\sqrt{\alpha^2+i\alpha s},\lambda-\frac{i}{N}\sqrt{\alpha^2-i\alpha s})$. A naive saddle point approximation leads to $r_{1,2}=1/2(i\lambda\pm 2\pi \rho)$ but substituting it back to eq.(\ref{SSym}) shows that the integrand vanishes at the saddle-point value. Finding a way to fully control the higher order expansion around  the saddle point and extract all relevant contributions remains a challenge. It is easy to see however that the result of the expansion will satisfy the rescaling property (\ref{Rescale}).

\subsection{Appendix B}
Our goal is to give a brief derivation of the explicit expressions for $\langle\overline{(\Re \mathbf{K}_{a,b})^2}\rangle_{GOE(N)}$ and $\langle\overline{(\Im \mathbf{K}_{a,b})^2}\rangle_{GOE(N)}$.

We start by rewriting $\mathbf{K}_{ab}$ in the basis of the eigenvectors of $\mathbf{H}_N$, namely:
$$
\mathbf{K}_{ab}=\sum_{n=1}^{N}\frac{\sum_{i,j=1}^{N}w_{a,i}(\mathbf{O})_{in}(\mathbf{O})_{nj}w_{b,j}}{(\lambda-\lambda_n)^2+\alpha^2/N^2}.
$$
It is easy to see that after performing the averaging over the Gaussian channel vectors we can write
\begin{equation}\label{K2}
\fl \lim_{N\rightarrow\infty}\langle\overline{(\Im \mathbf{K}_{a,b})^2}\rangle_{GOE(N)}=-\lim_{N\rightarrow\infty}\frac{\alpha}{2N}\frac{d}{d\alpha}\Big(\frac{1}{\alpha}\Im\Big\langle\Tr\Big\{\frac{1}{(\lambda-i\alpha/N){\bf 1}_N-\mathbf{H}_N}\Big\}\Big\rangle_{GOE(N)}\Big)
\end{equation}
and:
\begin{equation}\label{K3}
\fl \lim_{N\rightarrow\infty}\langle\overline{(\Re \mathbf{K}_{a,b})^2}\rangle_{GOE(N)}=\lim_{N\rightarrow\infty}\frac{1}{2N\alpha}\frac{d}{d\alpha}\Big(\alpha\Im\Big\langle\Tr\Big\{\frac{1}{(\lambda-i\alpha/N){\bf 1}_N-\mathbf{H}_N}\Big\}\Big\rangle_{GOE(N)}\Big)
\end{equation}
The traces above in the limit of $N\rightarrow \infty$ can be written in terms of the Stieltjes transform of the semicircle density:
$$
\lim_{N\rightarrow \infty}\Big\langle\frac{1}{N}\Tr\frac{1}{\mathbf{H}_N-z}\Big\rangle_{GOE(N)}=\frac{1}{2\pi J}\int_{-2}^{2}\frac{\sqrt{4-x^2}}{x-z/J}dx
$$
$$
=\frac{\lambda-i\alpha/N}{2J^2}\Big(-1+\sqrt{1+\frac{4J^2}{(\alpha/N+i\lambda)^2}}\Big)
$$
The imaginary part can be extracted by observing that $\sqrt{a+ib}=x+iy$ where $x=1/\sqrt{2}\sqrt{\sqrt{a^2+b^2}+a}$ and $y=\mbox{sign}(b)/\sqrt{2}\sqrt{\sqrt{a^2+b^2}-a}$. Performing the derivatives in $\alpha$ leads to:
$$
\langle\overline{(\Re \mathbf{K}_{a,b})^2}\rangle_{GOE(N)}=\frac{\sqrt{4J^2-\lambda^2}}{4J^2\alpha},
$$
and the same result holds for $\langle\overline{(\Im \mathbf{K}_{a,b})^2}\rangle_{GOE(N)}$.

\end{document}